\documentclass[sigconf]{acmart} 
\AtBeginDocument{%
  }

\setcopyright{acmlicensed} 
\copyrightyear{2026} 
\acmYear{2026} 
\acmDOI{XXXXXXX.XXXXXXX} 
\acmConference[CCS '26]{Make sure to enter the correct
  conference title from your rights confirmation email}{Nov 15--19,
  2026}{Hague, Netherlands}  
\acmISBN{978-1-4503-XXXX-X/2018/06}  

\usepackage{tikz}
\usepackage{amsmath}
\usepackage{enumitem}
\setlist{nolistsep}
\usepackage{booktabs}
\usepackage{xspace}
\usepackage{multirow}
\usepackage{tabularx}
\usepackage{xcolor}
\usepackage{wasysym}
\usepackage{float}
\usepackage[utf8]{inputenc}
\usepackage{textgreek}
\usepackage[normalem]{ulem}
\usepackage{tikz}
\usepackage{pgfplots}
\pgfplotsset{compat=1.18}
\usepackage{sansmath} 

\hypersetup{
  colorlinks,
  linkcolor={green!100!black},
  citecolor={red!80!black},
  urlcolor={blue!70!black}
}

\newcommand{\addd}[1]{#1}

\newcommand{\removee}[1]{}

\newcommand{\add}[1]{#1}

\newcommand{\remove}[1]{}

\begin{document}



\title{``People can change, and patterns can be broken'': Contextualizing Tradeoffs in Automated Decision-Making Systems} 





 \author{Rabeya Bosri}
 \affiliation{%
   \institution{University of Alberta}
   \country{Canada}
 }

 \author{Anna Harbluk Lorimer}
 \affiliation{%
   \institution{University of Chicago}
   \country{USA}
 }

  \author{Afrida Hossain}
 \affiliation{%
   \institution{University of Alberta}
   \country{Canada}
 }

  \author{Vasisht Duddu}
 \affiliation{%
   \institution{University of Waterloo}
   \country{Canada}
 }

  \author{Bailey Kacsmar}
 \affiliation{%
   \institution{University of Alberta}
   \country{Canada}
 }


\begin{abstract} 

Automated decision-making (ADM) systems are increasingly deployed in domains such as mortgage lending, prison sentencing, health insurance coverage, and hiring. Designing a responsible ADM system in such high-stakes domains requires ensuring privacy protection, fairness across demographic groups, and robustness against adversarial manipulation. 
However, prioritizing one of these objectives comes at the cost of another, forcing a choice as to which tradeoff to accept in a deployment. 
These tradeoffs \add{explicitly or implicitly} impact the life, safety, and fundamental rights of \add{the people in a society}\remove{those to whom the ADM is applied}, and thus, the perceptions and priorities of this population are needed before we can produce appropriate solutions. To this end, we conducted a quasi-experimental study (N = 777) in which participants evaluated four decision-making scenarios with controlled tradeoffs. Participants significantly preferred human decision-making (HDM) over ADM in three of four scenarios, emphasizing the value of human judgment, contextual understanding, and the ability to incorporate non-quantifiable factors. Furthermore, in terms of tradeoffs, our findings not only show that participants' preferences are highly context-dependent, but also that their perception of a specific objective, fairness, extends beyond formal definitions. Participants interpret fairness through multiple lenses, including privacy risks and susceptibility to manipulation, and view unfair or manipulated outcomes as failures of accuracy.
Overall, our findings highlight the importance of context-aware and human-centered approaches when designing and governing ADM systems in high-stakes situations. Rather than purely technical objectives, it is essential to evaluate ADM systems based on how their tradeoffs align with specific expectations within a given domain, as well as with societal values and perceptions of harm and fairness.

\end{abstract}

\ccsdesc[500]{Security and privacy~Usability in security and privacy}
\ccsdesc[500]{Security and privacy~Human and societal aspects of security and privacy}


\keywords{Differential Privacy, Group Fairness, Adversarial Manipulation, User Study} 



\maketitle

\section{Introduction}
Scenarios in which humans were primarily responsible for decision-making are increasingly being delegated to automated decision-making (ADM) systems. 
These ADM systems rely on patterns learned from data, primarily historical data,  to make decisions in new contexts \cite{mitchell1980need}.
When such ADM systems are applied to high-stakes contexts, where decisions affect an individual’s health, safety, or fundamental rights, the consequences can be severe.
%
Over the last two decades we have seen multiple instances in which the use of ADM systems has raised concerns about unfair outcomes for certain demographic groups~\cite{reutersAmazonBias2018,applebiasNews,criminal}, concerns about the privacy of the training data~\cite{privacySurvey}, concerns for adversarial tampering with the system~\cite{trades,madry2018towards}.
As these systems are increasingly deployed in sensitive domains such as finance, recidivism, and healthcare~\cite{ trinh2024algorithmic,DresselJulia2018Tafa}, the research community has developed techniques for addressing each of these concerns in isolation, but
addressing these concerns in combination has proved difficult. 

For example, in the case of privacy, Differential Privacy (DP) has been employed for ADM that uses machine learning (ML)~\cite{dwork2006differential}. However, deploying DP can disproportionately harm an ML model's performance for minority groups, introducing a tradeoff between privacy guarantees and fairness across demographic subpopulations for ADM systems that use machine learning~\cite{dpaccdisp}. 
Similarly, techniques that mitigate certain types of tampering
also entail unintended risks, such as increased vulnerability to privacy attacks, further illustrating the complex interplay among the desired attributes of ADM systems \cite{xu2018fairgan}.
These fundamental tensions must be addressed to deploy ADM systems that meet user expectations. However, current research has not delved into how users perceive these tensions and how they prioritize tradeoffs. 

In summary, the deployability of ADM systems is intertwined with privacy, fairness, and adversarial tampering concerns, and efforts to address one concern can come at the cost of another. 
Since mitigating one concern may exacerbate another, the question arises as to which objectives should be prioritized and who should make these decisions. Among all the stakeholders involved in ADM systems, \remove{the people ADM systems make decisions about represent the largest stakeholder group.}\add{the people in a society who will be implicitly or explicitly affected by the ADM decisions represent the largest stakeholder group.} They are \remove{directly} affected by the outcomes of ADM systems, making their perspectives crucial for guiding their design and deployment. 

Therefore, we explore \remove{users'}\add{peoples'} acceptability and perceptions of HDM and ADM in high-stakes decision-making scenarios, and how such \remove{users}\add{stakeholder groups} perceive and prioritize tradeoffs among privacy, fairness, and security in these scenarios. To this end, we ask and answer the following research questions. \\
\textbf{RQ1:} What is the users' acceptability of HDM and ADM in high-stakes scenarios, and how does their acceptability differ across scenarios? \emph{(Sect.~\ref{sect:quantADMHDM})}. \\ 
\textbf{RQ2:} What considerations, concerns, and values shape users' acceptability of HDM and ADM in high-stakes scenarios? \emph{(Sect.~\ref{sect:qualHDMADM})}.\\
\textbf{RQ3:} What is the users’ preference of tradeoffs among privacy, fairness, and security in high-stakes ADM scenarios? \emph{(Sect. ~\ref{tradeoffs_quan} and \ref{tradeoff_across_scenario})}.
\textbf{RQ4:} What considerations, concerns, and values shape users' perception of prioritizing the tradeoffs in high-stakes ADM scenarios? \emph{(Sect.~\ref{sect:considerationstradeoffs})}.

In brief, while participants appreciate ADM for its efficiency and perceived objectivity in structured tasks, they value HDM for its context-appropriate judgments and ability to consider aspects beyond mere data. Participants raised concerns about potential biases in both HDM and ADM, as well as technical and human limitations.
Participants prioritized tradeoffs contextually, demonstrating a willingness to accept biases, compromise on privacy, or risk adversarial manipulation to achieve desired outcomes. Their decisions were influenced by the stakes involved, the sensitivity of the information, and the potential consequences of accepting risks, as they weighed costs against benefits.





\begin{table}[htb]
\caption{Tradeoffs are emphasized with
a \textcolor{red}{$\Diamondblack$}\xspace to denote that using the defense makes that negative outcome worse. Unexplored cases are indicated with \textcolor{blue}{$\LEFTcircle $}\xspace, and where preventing that outcome is the goal of the defense is indicated $N/A$.}
\label{tab:unintended}
\footnotesize
\begin{center}
\begin{tabular}{ lccc } 
\bottomrule

\toprule
~ & \multicolumn{3}{c}{\textbf{Defended Objective}} \\
\textbf{Outcomes} & {Security} & {Privacy} & {Fairness}\\ 
\midrule
{Utility Loss} & \textcolor{red}{$\Diamondblack$} \cite{robustVacc,robustVacc2} & \textcolor{red}{$\Diamondblack$} \cite{dpsgd,dpVacc} & \textcolor{red}{$\Diamondblack$} \cite{accfairtradeoff,rodolfa2021empirical}\\ 


{Evasion} & $N/A$ & \textcolor{red}{$\Diamondblack$} \cite{robustVDP1,robustVDP2}& \textcolor{red}{$\Diamondblack$} \cite{tran2022fairness}\\ 

{Poisoning} &  $N/A$ & \textcolor{red}{$\Diamondblack$} \cite{poisonDP,hong2020effectiveness}& \textcolor{red}{$\Diamondblack$} \cite{adversarialBias,Mehrabi2021,poisoningFairML,poisonFair}\\ 

{Membership Inference} & \textcolor{red}{$\Diamondblack$} \cite{MIARobustness,Hayes2020TradeoffsBM}& $N/A$   & \textcolor{red}{$\Diamondblack$} \cite{miafairness,tian2023fd}\\ 


{Attribute Inference} & \textcolor{blue}{$\LEFTcircle $} &    $N/A$ & $N/A$ \\  

{Distribution Inference} & \textcolor{red}{$\Diamondblack$} \cite{suriSatml}& $N/A$  & \textcolor{red}{$\Diamondblack$} \cite{suriSatml}\\ 

{Data Reconstruction} & \textcolor{red}{$\Diamondblack$} \cite{robust2019private,privacyFLrobust}&   $N/A$ &  $N/A$ \\  

{Discriminatory Behavior} & \textcolor{red}{$\Diamondblack$} \cite{robustOddFair,hu2022understanding,ma2022on}  & \textcolor{red}{$\Diamondblack$} \cite{dpaccdisp,GongMIAUnfair,dispvuln} & $N/A$\\  
\bottomrule

\toprule
\end{tabular}
\end{center}
\end{table}
\section{Background}\label{sec:background}

\paragraph{Automated decision making systems}

In this paper, we refer to all systems that automate decision-making through machine learning (ML), artificial intelligence (AI), or data-driven analysis as ADM systems. ADM systems have been deployed to replace or augment HDM across varied real-world high-stakes scenarios. ADM systems are used to automatically determine credit limits~\cite{applebiasNews} and automatic health risk assessments to determine coverage~\cite{48431}. Meanwhile, in hiring, automated resume screening and hiring systems have been deployed to filter, score, and prioritize applicants~\cite{bogen2018help, raghavan2020mitigating}, ultimately influencing which applicants get hired. Finally, 
ADM systems have also been utilized for automated risk assessment in criminal justice systems worldwide~\cite{OswaldMarion2018Arap, ShepherdStephaneM_2014VRAi}.  
Notably these systems are deployed with the intention to improve decision  making processes in settings that have real, material impacts on people's lives. Thus, while we can appreciate their potential to improve these systems, we must also consider their weaknesses. 


\paragraph{ADM risks}
Our selected classes of vulnerabilities and attacks have been selected because their associated defenses pose tradeoffs for the valued objectives of privacy, fairness, and tamper-resistance. 
We refer to attacks that adversarially impair the ADM system as \emph{security attacks}. We emphasize, that while we use this term for brevity, we do not consider all security attacks. We limit security attacks to model manipulation via evasion or via poisoning. 
A malicious client can tamper with a model using \emph{evasion} by manipulating inputs to the model to force a misclassification, such as to force approval of loan that would not be approved otherwise~\cite{madry2018towards}. 
Similarly, a model can be manipulated by adding tampered data records to its training data, leading it to perform poorly on some data records, or by manipulating its predictions during inference for specific inputs~\cite{gu2017badnets}. This technique is known as \emph{poisoning} and could enable a hiring manager to manipulate an ADM such that it prioritizes scoring certain applicants higher than others.



We refer to attacks that impair the privacy of the training data as \emph{privacy risks}. 
We place privacy attacks that target the queries to a model outside the scope of this work. Our set of relevant privacy attacks are {membership inference}, {attribute inference}, {distribution inference}, and {data reconstruction}. 
In a \emph{membership inference} attack, an adversary's goal is to determine whether a target data record was used to train the model~\cite{shokri2017membership}. For example, a successful membership inference attack could reveal if a given patient's data was used to train an ADM that determines if a patient has depression, therefore revealing the patient's diagnosis. 
An \emph{attribute inference} attack refers to when an adversary successfully determines the values of a sensitive attribute associated with the training data set~\cite{attinf}. For example, an attacker could learn that all of the data in the dataset came from women.
In the case of a \emph{distribution inference} attack, the adversary's goal is to determine the rate of representation of a property across the dataset~\cite{propinf}. For instance, they may wish to learn the ratio of men to women as their distribution inference attack. 
Finally, in a \emph{data reconstruction} attack, an adversary is successful if they are able to reconstruct training data from their access to the model, typically via querying the model~\cite{carlini2023extracting,nasr2023scalable}.


Finally, we use \emph{fairness risks} as the term to broadly encompass different discriminatory outcomes associated with a model~\cite{fairness_def}. For example, a creditworthiness algorithm and how it was used to determine credit limits for credit card applicants. One such system made determinations that women found unfair because they received lower credit limits than men with similar financial histories~\cite{applebiasNews}.



%

\paragraph{Tradeoffs among defenses and risks}
We use \emph{tradeoffs} to refer to when techniques to preserve one desirable property, namely to mitigate a security, privacy, or fairness risk, do so at the detriment of another desirable property. For example, a privacy risk mitigation may hamper the ADM system's accuracy in its decision making~\cite{cummings2019compatibility}.
Table~\ref{tab:unintended} provides an overview of the tradeoffs and 
indicates which defense techniques for these objectives have an associated negative outcome~\cite{duddu2023sok}. 
Real-world deployment of ADM requires practitioners to account for these tradeoffs and find the best strategy to balance them, ensuring compliance with the law. Otherwise, the institution might face legal consequences.

\paragraph{Law and policy} 
New laws and policies are being developed to ensure that ADM systems are safe, fair, resilient against adversarial manipulation, maintain public trust, and comply with existing privacy regulations. For example, the United States is implementing AI regulations at the state level, including Colorado, Texas, and Utah~\cite{coloradoAI, uthaAI, texasAI}, as well as domain-specific regulations, such as the Illinois Artificial Intelligence Video Interview Act for hiring~\cite{illinoisJob}. Beyond the United States, the European Union (EU) has established a regulation regarding semi or fully ADM systems \cite{euai}. The EU Act identifies several high-risk applications, including medical risk assessment (Annex III, point 5(c)), candidate screening for jobs (Annex III, point 4(a)), criminal risk assessment (Annex III, point 6(a)), and financial decision-making (Annex III, point 5) \cite{euai}. 
When practitioners deploy automation in high-risk domains, they need to ensure compliance with legal and regulatory requirements. Therefore, it is important to understand the perspectives of the people ADM systems will directly impact, so that ADM design choices meet not only technical metrics, but also people's priorities. 

\section{Related Work}\label{sec:related}
Perceptions of privacy, security and fairness have each been considered in isolation in the literature. 
While we bring all of these objectives together in our study, we first highlight users' perceptions of the objectives in isolation.

In ADM systems, privacy risks arise at multiple levels, including risks to individuals in the training data and risks related to the disclosure of aggregate data properties~\cite{shokri2017membership, suriSatml}.
Protections against these risks include techniques such as differential privacy, secure multiparty computation, and homomorphic encryption, which, when applied, are each known to have some impact on how users understand and evaluate a system~\cite{DPuserStudy, kacsmar2022caring, distler2020making, dechand2019encryption}. 
The trust and acceptability of these privacy protection techniques depends on both their formal guarantees as well as on the application context, such as the domain being healthcare, finance, or surveillance~\cite{mehdy2021privacy, kacsmar2022caring, zheng2018user}. 
In addition to the technical guarantees and application context, user perceptions of privacy also vary across cultural and demographic contexts, reflecting not only individual preferences, but also broader social norms~\cite{abokhodair2016privacy, naveed2022ask, li2022cultural, cai2023understand, kablo2025privaci, haider2025effect}.

Fairness in ADM systems is similarly highly context-dependent. 
In low-stakes domains such as social media, users evaluate fairness in terms of visibility and opportunity allocation~\cite{choi2022s, ma2022m}. In contrast, in high-stakes settings like criminal justice, fairness is judged by focusing on whether decisions are made without bias and are proportional to the offenses committed~\cite{binns2018s, Harrison20}. Broadly, fairness perceptions are shaped by both algorithmic factors, such as accuracy, and human factors, including moral intuitions and subjective beliefs, as well as sociodemographic characteristics~\cite{wang2020factors, grgic2018human, starke2022fairness}. Additionally, users showed disagreement on which fairness metrics are appropriate, such as demographic parity, equal opportunity, and equalized odds~\cite{Srivastava19, grgic2018human, fairness_def}. While algorithmic fairness metrics capture different notions of fairness, users tend to prioritize accuracy, particularly in high-stakes contexts such as medical decision-making \cite{Srivastava19}, and prefer predictive parity when sensitive attributes are involved~\cite{fairness_def}.
Overall, this suggests that fairness is not a fixed property of a system but a socially constructed perception that can vary based on scenarios and stakeholders~\cite{starke2022fairness, abu2020contextual, li2022contextualized, parziale2025contextual}.

Finally, beyond fairness and privacy we have the notion of resistance to adversarial manipulation. Unlike the prior two objectives which have plentiful literature on perceptions, prior work on adversarial manipulation is largely algorithmic and related to technical measures, with limited attention to the perceptions of those who are negatively impacted by failures to prevent such manipulations. 
Existing studies on resistance to adversarial manipulation, or robustness, perceptions are largely limited to domains such as recommendation systems and autonomous vehicles~\cite{neyazi2023understanding, wei2025understanding}.

Overall, while existing studies focused on users' perceptions of an objective in isolation, deployed systems must consider the objectives collectively, since achieving one comes at the cost of losing another. Further, it is important to understand whether peoples' priorities are the same across different ADM systems or whether they dynamically re-prioritize competing values across scenarios before there can be a clear process on how to select among these objectives when producing a real system for deployment or when designing techniques with different tradeoffs in research.

\section{Methods}\label{sec:methods}
We employ a quasi-experimental design to investigate how \remove{user} \add{participants'} perception and acceptance of HDM and ADM differ across four scenarios. For these scenarios, we also explore how \remove{users}\add{participants} prioritize and perceive different tradeoffs.
See Appendix~\ref{appendix_ethics} and ~\ref{appendix_positionality} for our ethical considerations and positionality statement.

\begin{table*}[htb]
\begin{center}
\caption{The following is an overview of our scenarios, where Company A makes determinations about the given scenario. Company B has made an ADM system to assess an aspect relevant to Company A's decisions.
}

\label{tab:classification_task}
\begin{tabular}{ l  l  l l }
\toprule
\textbf{Scenario Domain} & \textbf{Company A} & \textbf{Company B's ADM} & \textbf{Sensitive Attribute} \\ 
 \midrule
 \textbf{Medical Insurance} & High/Low insurance premium & Risk of heart disease & Mental health issues\\  
 \textbf{Job Hiring} & High/Low chance of hiring & Risk of bad fit hire  & Criminal background\\
 \textbf{Prison} & Long/Short prison sentence & Risk of re-offending & Wealthy inmates \\
 \textbf{Mortgage} & High/Low interest rate & Risk of defaulting  &  Minority subgroup\\
 \bottomrule
\end{tabular}
\end{center}
\end{table*}

\subsection{Study Components}
\paragraph{Overview of the study}
As an overview, we present how a participant would experience the flow of our study (see study instrument in Appendix~\ref{appendix_questions}). 
First, participants were randomly assigned to one of four application domains, which we term our scenarios. 
Table~\ref{tab:classification_task} overviews our scenarios. Participants were asked to indicate their preference for HDM and ADM in that scenario on a scale from completely unacceptable to completely acceptable and explain in a free-form response. 
Second, participants were asked to give their preference in that scenario for one set of our two sets of tradeoffs by indicating a selection and providing a free-form explanation. Finally, all participants answered the same demographic questions. 




\paragraph{Scenarios and tradeoffs}
For our selected scenarios, we focus on cases where deploying an ADM would affect people's lives, safety, and fundamental rights. 
We selected our scenarios based on the real-world ADM applications in Section~\ref{sec:background} and the high-risk settings in the EU AI Act. 
The resulting scenarios are \emph{medical insurance, job screening, prison sentencing, and mortgage approval}. 

Among the known tradeoffs from Table~\ref{tab:unintended}, we select those where the risk of a negative outcome increases when the defense is deployed. 
Note there are positive relationships between 
objectives and
a defense. For example, defending privacy uses techniques that directly counteract membership inference, attribute inference, distribution inference, and data reconstruction. 
The tradeoffs in our study, therefore, include the following eight combinations. First, defending fairness versus each of utility loss, membership inference, distribution inference, and security risk. Second, defending security risk versus each of utility loss, membership inference, distribution inference, and data reconstruction. These correspond to our two sets, with each participant receiving only one of the sets. To effectively communicate the tradeoffs to participants, we have used terms that are participant-friendly.
The term accuracy is presented as a correct result. We describe bias as treating one group of people more favorably than another.
In the case of fairness and accuracy, the participants received a set of options, where being favored as a brunette is the proxy for bias. Choices would include ``{The [ADM] strongly favors brunettes; otherwise, it provides the correct chance of being hired}'', ``{The [ADM] moderately favors brunettes, and it provides
slightly incorrect chance of being hired}'', ``{The [ADM] slightly favors brunettes, and it provides a moderately incorrect chance of being hired}'', and ``{The [ADM] does not favor brunettes, and it provides a highly incorrect chance of being hired}'.
In the case of membership inference, we refer to an employee's ability to determine whether an individual's sensitive data was used in training the ADM. Distribution inference indicates an employee's capacity to estimate the number of women in the training dataset. We define adversarial manipulation as an adversary modifying the ADM to produce incorrect results rather than true ones. Finally, data reconstruction is the process of recovering sensitive information that was part of the ADM system's training data. 

 \paragraph{Pilot}
We piloted the survey with individuals from both computing science and non-computing science backgrounds to assess how well the scenarios were understood before the final survey. In the first iteration of the survey, we used a five-point scale to indicate a strict preference for protecting one risk over the other, a slight preference, or no preference. Based on feedback, we revised the format to remove the neutral option (no preference) and present explicit paired tradeoffs such as (high, none) and (medium, low), which encouraged participants to make a choice. We then updated our scale and provided additional text and context for the participants. 


\subsection{Participants}
Participants were recruited via Prolific from April 2025 to August 2025. \remove{,resulting in $777$ completed responses}\add{To mitigate bot interference, Prolific deploys extensive bot and AI-agent detection tools to prevent non-human  responses~\cite{prolific}.} 
\add{To further ensure data quality, two researchers manually reviewed all free-form responses and discarded responses that were disconnected from the questions. We discarded 10 responses in total, resulting in $777$ completed responses.} Participants spent an average of $15$ minutes on our study and received $\$3$ USD. Participants were permitted to withdraw from the study at any time and could skip any question they chose.

All participants were located in the United States, \add{ and we identified them as \emph{experiential experts}~\cite{young2019toward}, people living the experience, or associated with someone living the experience. We recruit participants with various demographic attributes and who are members of a society that may be explicitly or implicitly affected by ADM in domains such as hiring, mortgages, insurance, and criminal justice. We do not know which of these they have explicitly or implicitly experienced in each setting in their lives.} Among the total $N = 777$ participants, 48\% indicated they were women, 51\% indicated they were men, and 1\% indicated they were non-binary.
Regarding minority group representation, 43\% indicated they belonged to a minority group, 54\% indicated they did not, and 3\% preferred not to answer.
Participants ages were distributed as follows with 13\% between 18-25 years old, 19\% between 26-35, 17\% between 36-45, 19\% between 46-55, 28\% between 56-65, and 4\% preferred not to disclose their age.
In terms of educational background, less than 1\% of participants had not completed high school or the equivalent, 10\% had a high school diploma or equivalent, 3\% had some non-university education beyond high school, 20\% had some college without obtaining a degree, 44\% earned a bachelor’s or associate degree, 17\% had a master’s degree, 6\% a doctoral degree, and 1\% preferred not to disclose their educational background. Regarding employment as ML practitioners, 86\% reported not working in the field, 12\% indicated they do, and 2\% preferred not to answer.

\begin{figure*}[t]
\centering
\caption{The distribution of preferences for HDM and ADM across all scenarios. Acceptability is measured on a five-point scale, and each segment corresponds to the proportion of participants who selected that level of acceptability. 
The sample size ($n<N$) is provided and for interpretability we include a mean acceptability score ($\mu$) and confidence interval (CI) for each condition.}
\noindent
\makebox[\textwidth][l]{
\begin{tikzpicture}
\begin{axis}[
    xbar stacked,
    width=.63\textwidth,
    height=4.4cm,
    xmin=0, xmax=100,
    bar width=1,
    xlabel={Proportion of $n$ Respondents (\%)},
    yticklabel style={
        text width=7.8cm, 
        align=right
    },
    ytick={8,7,6,5,4,3,2,1},
    yticklabels={
    {Prison - HDM ($n=195,\ \mu=3.12,\ 95\%\ \mathrm{CI}\ [2.96,\,3.29]$)},
   {Prison - ADM ($n=195,\ \mu=2.79,\ 95\%\ \mathrm{CI}\ [2.63,\,2.93]$)},
   {Mortgage - HDM ($n=197,\ \mu=3.95,\ 95\%\ \mathrm{CI}\ [3.80,\,4.10]$)},
   {Mortgage - ADM ($n=197,\ \mu=3.15,\ 95\%\ \mathrm{CI}\ [2.98,\,3.32]$)},
    {Job - HDM ($n=192,\ \mu=3.90,\ 95\%\ \mathrm{CI}\ [3.75,\,4.04]$)},
    {Job - ADM ($n=192,\ \mu=3.05,\ 95\%\ \mathrm{CI}\ [2.87,\,3.23]$)},
    {Insurance - HDM ($n=192,\ \mu=3.12,\ 95\%\ \mathrm{CI}\ [2.93,\,3.32]$)},
    {Insurance - ADM ($n=192,\ \mu=3.12,\ 95\%\ \mathrm{CI}\ [2.93,\,3.32]$)},
    },
    ymin=0.6, ymax=8.4,
    nodes near coords,
    every node near coord/.append style={
        font=\footnotesize,
        color=black
    },
    nodes near coords align={center},
    legend style={
        font=\footnotesize,
        draw=none,
        at={(0.5,1.02)},
        xshift=-3.5cm, 
        anchor=south,
        legend columns=5,
        /tikz/every even column/.append style={column sep=0.25cm}
    }
]
\addplot+[xbar, fill=blue!40!cyan, draw=white]
coordinates {
    (10.3,8) [10.3]
    (16.4,7) [16.4]
    (1.6,6) [1.6]
    (12.7,5) [12.7]
    (2.6,4) [2.6]
    (13,3) [13]
    (16.7,2) [16.7]
    (12.6,1) [12.6]
};

\addplot+[xbar, fill=orange!75, draw=white]
coordinates {
    (23.1,8) [23.1]
    (26.7,7) [26.7]
    (10.7,6) [10.7]
    (17.8,5) [17.8]
    (8.3,4) [8.3]
    (25,3) [25]
    (21.4,2) [21.4]
    (20.9,1) [20.9]
};

\addplot+[xbar, fill=gray!25, draw=white]
coordinates {
    (21,8) [21]
    (24,7) [24]
    (15.2,6) [15.2]
    (23.4,5) [23.4]
    (15.1,4) [15.1]
    (17.7,3) [17.7]
    (11.5,2) [11.5]
    (18.3,1) [18.3]
};

\addplot+[xbar, fill=green!35, draw=white]
coordinates {
    (35.4,8) [35.4]
    (26.7,7) [26.7]
    (36.5,6) [36.5]
    (34.5,5) [34.5]
    (44.8,4) [44.8]
    (32.3,3) [32.3]
    (33.9,2) [33.9]
    (38.7,1) [38.7]
};

\addplot+[xbar, fill=magenta!30, draw=white]
coordinates {
    (10.2,8) [10.2]
    (6.2,7) [6.2]
    (36,6) [36]
    (11.6,5) [11.6]
    (29.2,4) [29.2]
    (12,3) [12]
    (16.5,2) [16.5]
    (9.4,1) [9.4]
};

\legend{
Completely Unacceptable, 
Somewhat Unacceptable, 
Neutral,
Somewhat Acceptable,
Completely Acceptable
}
\end{axis}
\end{tikzpicture}
}
\vspace{-0.7cm}
\label{fig:accepthumanai_revision}
\end{figure*}
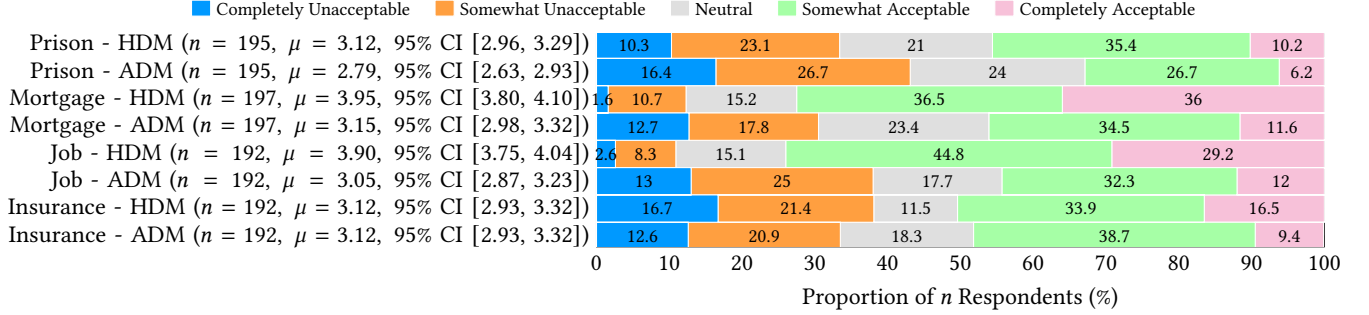
\subsection{Analysis Methods}

For our quantitative analysis, we used the Wilcoxon signed-rank test, a non-parametric test for paired data, as each participant evaluated both HDM and ADM. Further, we report two effect sizes, the rank-biserial correlation (\(r_{rb}\)) and the standardized effect size \(r\) (derived from the test statistic \(Z\)). 
To examine participants' preferences across the four scenarios, we used the Kruskal–Wallis test and 
when it indicated a significant difference, we performed Dunn’s post hoc tests with Bonferroni correction to identify pairwise differences. For significant pairs, we used the Mann–Whitney U test to determine the direction of the effect, such as which condition yields higher acceptability ratings for HDM versus ADM in a given scenario. Collectively, the above analysis is used to address RQ1. 
\remove{We report descriptive statistics for our categorical data corresponding to our tradeoff pairs.}
\add{We conduct the Chi-Square test on our categorical data and report descriptive statistics corresponding to our tradeoff pairs.} Within each response set, two options prioritize one objective, and the remaining two prioritize the alternative objective in the tradeoffs. Accordingly, we summarize trends by considering the proportion of participants selecting options that prioritize each objective which we use to address RQ3. 


Finally, we conducted qualitative content analysis over the free-form responses to address RQ2 and RQ4. In the first coding cycle, we applied descriptive coding~\cite{alma991031555799709116} and maintained an evolving codebook. 
Two researchers independently \add{conducted the initial coding on the same data}\remove{coded the data} using the codebook and \remove{communicating updates to}\add{updated} it as needed. This allowed us to find codes inductively from participants’ responses. After coding each question, \remove{we compared assigned codes}\add{one researcher identified any response-level discrepancies in the coded data} \remove{at the response level for the two coders} to identify inconsistencies, which were resolved through discussion to reach consensus.
\add{ When the two researchers could not reach a consensus, a third researcher was included.}
In the second coding cycle, we employed focused coding~\cite{alma991031555799709116} to refine and integrate the initial codes. Finally, we collaboratively grouped our codes into broader categories based on semantic similarity and conceptual relationships, and derived higher-level themes by identifying patterns and connections across the categories. 

\subsection{Limitations}
Our scenarios do not encompass all possible types of high-risk ADM. Rather, the ADM systems selected were chosen for their potential to have significant consequences in end-users’ lives.
Our participants are WEIRD (Western, Educated, Industrialized, Rich, and Democratic)~\cite{schulz2018origins}. Consequently, our participant sample is not representative of the global population. The study relies on self-reported perceptions, which can be influenced by social desirability bias~\cite{redmiles2018asking}, misunderstandings of the scenarios, or individual differences in interpreting the questions.
Further, key concepts such as fairness were provided to participants with examples which may not perfectly represent all aspects of the underlying concepts. However, our pilot with individuals from both relevant computing science backgrounds and from non-computing backgrounds aided the representations efficacy for the scenarios. \add{Additionally, in the context of describing bias, we deliberately chose \emph{“favoring”}, over wording with negative connotations, to limit social desirability bias to the extent possible. Alternative framings may elicit different responses.} 
Finally, participants evaluated hypothetical scenarios,
which may not fully capture the complexity or emotional weight associated with actual high-risk ADM cases. Ultimately, the tradeoffs presented in the survey may simplify the complex technical and ethical considerations involved in real-world ADM.
\section{Acceptability Results for HDM and ADM}\label{sect:quantADMHDM}

Overall, participants responses tended to be more positive towards the HDM than to the ADM when considering the scenario prior to considering any defense tradeoffs. 
To provide intuitive insight into this trend, we present descriptive statistics before reporting our statistical test results within and across scenarios. See Figure~\ref{fig:accepthumanai_revision} for the participants’ acceptability responses. \addd{We also provide a graph of the mean score and 95\% confidence intervals as Figure ~\ref{fig:mean_ci} in the Appendix \ref{quantitative_human_AI}.}

We observed a higher proportions of participants select ``Somewhat Acceptable'' to ``Completely Acceptable,'' for HDM than for ADM prior to applying a defense. 
For example, HDM received positive acceptability ratings from 45.7\% of participants in the Prison sentencing scenario, 72.5\% in the Mortgage scenario, 75.8\% in the Job applications scenario, and 49\% in the Insurance scenario.
In comparison, ADM received positive acceptability ratings from 32.9\% of participants in the Prison scenario, 46.2\% in the Mortgage scenario, 44.3\% in the Job scenario, and 48\% in the Insurance scenario. 



For acceptability within the scenarios, we use the related-samples Wilcoxon Signed-Rank test with a significance threshold of \(\alpha = 0.05\) which showed a significant preference for HDM over ADM in all our scenarios except Insurance. In the Mortgage and Job scenarios, participants showed statistically significant preferences for HDM, with large effect sizes: $r_{rb}$ = $.59$ and $.57$ respectively, and $r$ = $-.49$ for both.  A preference for HDM was also observed in the Prison scenario, with small effect sizes: $r_{rb} = .17$ and $r = -.21$.  

\paragraph{HDM across scenarios}
Using a Kruskal-Wallis test revealed that participants' acceptability of HDM varies significantly when considering the different scenarios with a test statistic of $79.359$ and $p<0.001$. Consequently, we conducted post hoc pairwise comparisons to identify the specific pairs with significant differences. The following pairs had significant differences: Prison - Job, Prison - Mortgage, Insurance - Job, and Insurance - Mortgage. To further explore these differences, the Mann-Whitney U tests revealed that HDM was rated significantly higher in the Job and Mortgage scenarios when compared with Prison and Insurance scenarios.

\paragraph{ADM across scenarios}
Similar to HDM, the Kruskal-Wallis test revealed a significant overall difference in the acceptability of ADM across the four scenarios, with a test statistic of $10.527$ and $p=0.015$. Follow-up post hoc pairwise comparisons showed that the differences arose when the Prison scenario was compared against the Job and Mortgage scenarios. The Mann–Whitney U test showed that ADM was viewed as significantly more acceptable in the Job and Mortgage scenarios compared to the Prison scenario. No other scenario pairs exhibited significant differences. 

\paragraph{Acceptability by demographics}
We examined whether decision-making preferences differed by the following participant attributes: gender, minority status, and ML practitioner status. For gender, we compared only men and women, as the number of participants from other gender groups was too small for our statistical analysis.

For HDM, we found no significant differences for any of these attributes. 
For ADM, we found a significant difference by minority status in the Prison scenario $(p = 0.045, U = 3576.5, z = -2.0)$. The non-minority group showed greater acceptability of ADM than the minority group, with a higher mean rank (100.99 vs. 85.35). No other scenarios showed significant differences by minority status.
We also found a significant difference by ML practitioner status in the Insurance scenario $(p = 0.011, U = 1824.5, z = -2.533)$. ML practitioners had a higher mean rank than non-practitioners (116.48 vs. 90.62), suggesting greater acceptance of ADM for this domain. 

\section{Findings for HDM and ADM Considerations}\label{sect:qualHDMADM}
From our qualitative analysis of free-form responses, we identify aspects that shape participants' acceptance of HDM and ADM. 
We present participants' considerations, concerns, and values here organized into advantages and disadvantages.

\subsection{Perceived Advantages of HDM}\label{human_better}
Participants highlighted several benefits of HDM, with
 specific reasoning reflecting the needs and expectations of the scenarios.

\paragraph{Participants value looking beyond the numbers to make informed decisions} 
For all four scenarios, participants valued HDM when 
\begin{quote}
    ``assessing factors that require a nuanced understanding of human behavior and individual circumstances \ldots that a purely data-driven approach might miss, leading to a more holistic and potentially fairer assessment for individuals with complex personal histories \ldots'' [P247, Prison].
\end{quote}
Participants also identified scenario-specific advantages, such as in hiring, ``\ldots there are soft skills and interpersonal skills that can only be determined by a human \ldots"[P586, Job]. Further, participants expressed the need for human judgment to recognize ``grey areas" in the Mortgage scenario where
   ``\ldots younger adults, immigrants, or gig workers, may simply not have long traditional credit history \ldots'' [P373, Mortgage].
In such situations, ``\dots [HDM] can override strict data-based decisions" [P473, Mortgage]. 
Further, humans can demonstrate compassion and consider ``how an individual may have gotten into a situation in the past that could make them appear high-risk, but they are now in a better position'' [P477, Mortgage]. 


\paragraph{Participants value the human experience that guides informed decision-making}
Personal narratives, social and cultural experiences, work experience, and discerning observations are mentioned as enhancing decision-making. For instance, in the Prison sentencing scenario, ``some employees may have insight into the risk of re-offending based on their personal knowledge of situations" [P174].
In the Job scenario, participants noted that ``employees \ldots have likely been in the same shoes as the applicant at one point, so they may be well suited to assess an applicant's quality'' [P294, Job]. 
Across all scenarios, participants emphasized the existence of aspects that require human evaluation. 
Specifically, when the decisions might significantly impact individuals' lives,
``\ldots should be evaluated by people \ldots''[P767, Prison].

In summary, participants expressed that HDM has critical flexibility, unlike rigid rule-based systems, and enabling nuanced understanding and context-appropriate decisions. 

\subsection{Perceived Disadvantages of HDM}
Participants identified consistent disadvantages of HDM across scenarios, though the manifestation of harm was scenario-dependent. 
 
\paragraph{Participants raised concerns that social and personal biases compromise the fairness of HDM}
Across all the scenarios, participants noted that  using HDM to assess risk can be detrimental, 
\begin{quote}
    ``when decisions are influenced by personal biases \ldots such as when an employee unintentionally favors or discriminates against particular groups based on race, gender, or appearance \ldots'' [P56, Mortgage]
\end{quote}
Social prejudices can have significant consequences. In the Prison scenario,
participants expressed concerns that individual life experiences can also impact HDM, such as a ``\ldots family member having been a victim of a crime \ldots'' [P626, Prison].
Participants' concerns extend beyond personal benefits to include pressure from managerial priorities. 
In the Mortgage and Insurance scenarios, participants expressed concern that due to ``\ldots pressure to meet certain quotas, their assessments could be inconsistent, unfair, and potentially discriminatory \ldots'' [P80, Insurance].
There were also concerns for affinity bias, such as ``\ldots an [HDM] unconsciously prefers candidates who share similar backgrounds, communication styles, or educational paths \ldots [P336, Job],'' which can lead to unfair hiring practices and reduce organizations' diversity

\paragraph{Participants are concerned about inaccurate judgment, resulting from human limitations}
Participants described different dynamics of human emotion, such as sympathy, anger, and moral outrage, and how each can affect decision-making across scenarios. 
In the Insurance and Mortgage scenario, ``\ldots the client could get a higher rate if the employee has a negative perception towards the customer" [P492, Mortgage] while in the Prison scenario, ``\ldots due to our sympathetic nature we might make the wrong decisions \ldots'' [P210].
Participants also raised the issue of how certain work conditions can affect employees' emotional well-being and impact decision-making.
Apart from the emotional influence, ``factors like the employee's mood on a particular day, their personal experiences, or even subjective interpretations of behavior could lead to inconsistencies and unfair outcomes." [P245, Mortgage].

\subsection{Perceived Benefits of ADM}\label{ai_better}
In all scenarios, we observed a similar perception of how ADM can be beneficial in decision-making. Overall, participants' responses in this regard were not specific, focusing on ADM's generic properties rather than on how it can benefit a particular scenario. Below, we discuss the themes we found regarding the benefits of ADM. 

\paragraph{Participants believe that the efficiency provided by ADM will positively impact scalability} 
Participants emphasized that ADM enhances the efficiency of organizational processes. ADM can save time, resources, and fairness compared to HDM, thus,
 ``...is beneficial when there is a large volume of applications to process quickly and consistently" [P497, Mortgage].
Overall, ADM is described as a tool that enhances performance through faster processing and handling large-scale data efficiently. 


\paragraph{Participants value ADM for its perceived objectivity in decision-making}
Participants perceived ADM that promotes fairness in decision-making, with one participant noting that ``[ADM] is not racist.  Everyone would be treated equally and fairly'' [P481, Mortgage]. Unlike HDM, ADM does not experience any emotion, thus ``it would take emotions out of the equation \ldots'' [P80, Mortgage]. Especially in the Prison scenario, ``it could be beneficial since there will not be room for negative emotions towards the prisoner, and will focus on the facts provided ''[P649, Prison]. 

\paragraph{Participants perceive that ADM systems make data-driven decisions, which are accurate and reliable}
Participants perceived ADM as reliable, ensuring that decisions are based solely on relevant facts related to the task and thereby enhancing stakeholder confidence in the outcomes, as one participant articulated in Job scenario, ``the [ADM] will select applicants based on their experience and predict that they will perform well when hired \ldots'' [P105, Job]. Participants also emphasized the ADM’s ability to detect subtle correlations or anomalies that humans may overlook, reinforcing perceptions of higher accuracy, as highlighted in the following participant quote in the Insurance scenario, ``\ldots it could also spot hidden patterns in my health that even a doctor might miss" [P386, Insurance]. 


\paragraph{Participants believe ADM systems are suitable for repetitive tasks which are structured and unambiguous}
ADM were viewed as well suited for straightforward tasks where decisions follow clearly defined requirements. For example, ``finding the bare minimum of the job requirements, like education'' [P509, Job]. Participants felt that ADM could efficiently and consistently handle these tasks because they do not demand ethical interpretation, contextual understanding, or sensitivity to individual circumstances. As a result, ADM was seen as an appropriate tool for automating screening processes in which fairness and accuracy depend primarily on applying fixed rules such as, ``when a crime is clear cut, well-documented and has a lot of strong evidence'' [P316, Prison].

\subsection{Perceived Disadvantages of ADM}
Below, we discuss the themes that we have found and describe how they produce disadvantages across all four ADM scenarios.

\paragraph{Participants are concerned that technical limitations of ADM compromise the fairness and reliability of its decisions}
Participants reported scenario-specific correlations that might negatively influence decisions, such as certain illnesses leading to higher insurance premiums, living in certain neighborhoods increasing the likelihood of recidivism, and past criminality predicting future hiring. 
Participants highlighted that ADM are likely to perpetuate historical norms of discrimination, as ``the [ADM] might learn to treat similar applicants as higher risk, thereby perpetuating discrimination \ldots'' [P257, Mortgage]; which mirrors mortgage outcomes known to include structural racism~\cite{lewis2024racial}.
Additionally, participants expressed that accuracy depends on the training dataset; for example, ``\ldots if the [ADM] was trained predominantly on data from middle-aged men, it might inaccurately assess risk in women or younger applicants \ldots'' [P219, Insurance]. In fact, participants noted that ADM might ``\ldots  automates discrimination while hiding behind 'neutral' algorithms, violating ethical and legal standards'' [P314, Job]. 

\paragraph{Participants expressed concern that the lack of emotional intelligence in ADM leads to unreliable decision-making} 
Participants 
expressed concern that ADM systems cannot interpret emotional, moral, or social nuances.
They emphasized contexts where judgments must reflect not only facts, but also the human aspects of behavior and change as even when records indicate a high risk of recidivism or non-payment,``People can change, and patterns can be broken'' [P604, Prison].


\paragraph{Lack of governance and ethical accountability undermines participant trust in ADM}
Participants raised concerns about the lack of oversight and transparency, as well as the potential for corruption or privacy violations since the ``[ADM] could completely backfire if the wrong information is used and nobody notices'' [P383, Prison]. Furthermore, it is  ``...difficult to understand why a particular risk assessment was given, hindering transparency and the ability to appeal an unfair decision'' [P245, Prison]. This aligns with concerns that, without proper governance, it is unclear who is accountable for errors or harm. 

\paragraph{Participants showed concern about deploying ADM in high-stakes and sensitive scenarios}
The sensitivity of the decisions and the data involved in high-stakes scenarios produces a negative perception of ADM. Decisions which impacted peoples lives and well-being were described as concerning settings for ADM. For example, participants mentioned that ``using [ADM] for human health is dangerous and always a bad idea'' [P120, Prison] and even stated that ``any final decision made by a machine is a bad decision'' [P746, Mortgage].


\section{Tradeoff Preference Within Scenarios}
\label{tradeoffs_quan}

\addd{We test whether participants significantly chose one option more than others using a chi-square test (details in Appendix~\ref{appendix_tradeoffs_within_scenarios}) and found the following significant results. In the fairness versus accuracy tradeoff, excluding the Job scenario, participants chose the accuracy-maximizing option. In the fairness versus MI and fairness versus DI tradeoffs, participants choose to prioritize fairness in the Prison and Job scenarios.  In the fairness versus security tradeoff, participants chose the highest level of security in the Prison and Mortgage scenarios. However, in the security versus accuracy tradeoff, excluding the Prison scenario, participants chose the accuracy-maximizing option. 
}

\addd{Below we present the participants' preference distribution within scenarios. Recall that we cluster minimizing the risks of membership and distribution inference as ``prioritizing privacy". All percentages are for the $n<N$ participants assigned to that scenario. The following results show that participants' preferences are dependent on the specific scenario they are presented with. }

\remove{We discuss the participants' distribution in this section, and all the figures showing participants' distributions are included in the Appendix~\ref{appendix_tradeoffs_quant}.} 



\begin{figure}[!htb] 
\vspace{-0.7cm}
\centering \caption{Preferences of fairness versus accuracy along with sample size ($n<N$) for each scenario.} 
\begin{tikzpicture} \begin{axis}[
    xbar stacked,
   width=0.8\columnwidth,
    height=3cm,
    xmin=0, xmax=100,
    bar width=1,
    xlabel={Proportion of Respondents (\%)},
    ytick={4,3,2,1},
    yticklabels={
        Prison ($n=99$),
        Mortgage ($n=97$),
        Job ($n=99$),
        Insurance ($n=97$)
    },
    xtick={0,20,40,60,80,100},
    tick label style={font=\footnotesize},
    label style={font=\footnotesize},
    ytick style={draw=none},
    axis x line*=bottom,
    axis y line*=left,
    ymin=0.6,
    ymax=4.4,
    nodes near coords,
    every node near coord/.append style={
        font=\footnotesize,
        color=black
    },
    nodes near coords align={center},
    legend style={
    font=\footnotesize,
        draw=none,
        at={(0.5,1.02)},
        anchor=south,
        legend columns=2,
        /tikz/every even column/.append style={
            column sep=0.2cm
        },
        row sep=.5pt
    }
]
\addplot+[xbar, fill=blue!40!cyan, draw=white] coordinates { (45.5,4) [45.5] (43.3,3) [43.3] (37.4,2) [37.4] (45.4,1) [45.4] }; 
\addplot+[xbar, fill=orange!75, draw=white] coordinates { (35.5,4) [35.5] (32,3) [32] (31.3,2) [31.3] (33,1) [33] }; 
\addplot+[xbar, fill=gray!25, draw=white] coordinates { (9,4) [9.1] (8.2,3) [8.2] (13.1,2) [13.1] (11.3,1) [11.3] }; 
\addplot+[xbar, fill=magenta!30, draw=white] coordinates { (10,4) [10.1] (16.5,3) [16.5] (18.2,2) [16.5] (10.3,1) [10.3] }; 

\legend{\shortstack[l]{High bias; No Inaccuracy},\shortstack[l]{Moderate bias; Slight Inaccuracy},\shortstack[l]{Slight bias; Moderate Inaccuracy},\shortstack[l]{No bias; High Inaccuracy}} 
\end{axis} \end{tikzpicture} 
\label{t1_fair_accu_revision}
\vspace{-0.8cm}
\end{figure}
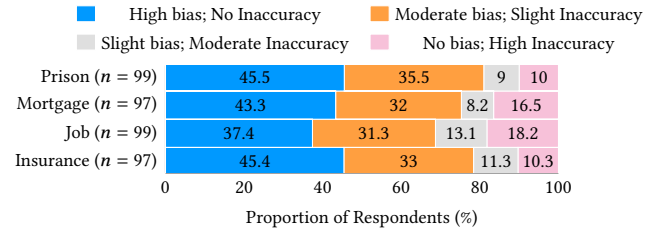

\subsection{Fairness versus Accuracy}\label{result_t1}
\removee{Excluding the Job scenario, participants significantly chose} 
\removee{ the accuracy-maximizing option over maximizing fairness; the distribution is in Figure~\ref{t1_fair_accu_revision}.}
\remove{Across the four scenarios, a larger proportion of participants selected accuracy-maximizing options over fairness, with varying degrees of intensity.} 
In the Prison scenario, 80\% of participants chose to prioritize accuracy, with 45\% accepting strong bias and 35\% accepting moderate bias as the consequence of their decision; \addd{the distribution is shown in Figure~\ref{t1_fair_accu_revision}}. 
In the Mortgage scenario, 75\% preferred accuracy-maximizing options, with 43\% accepting high bias and 32\% accepting moderate bias as the consequence. The same trend appeared for the Insurance scenario where 78\% opted for accuracy-maximizing choices, with the consequences split as 45\% for the strong-bias option and 33\% for the moderate-bias option. In the Job scenario, 68\% chose the accuracy-favoring options, and for the consequences, 37\% accepted high bias while 31\% accepted moderate bias. 



\subsection{Fairness versus Membership Inference}\label{result_t2}
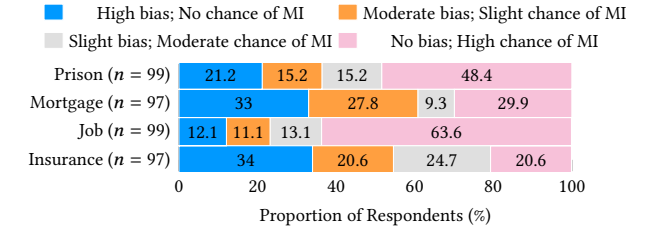
\begin{figure}[htbp]
\vspace{-.5cm}
\centering
\caption{Preferences of fairness versus membership inference (MI) with sample size ($n<N$) for each scenario.}
\vspace{0.8cm}
\noindent
\makebox[\textwidth][l]{
\begin{tikzpicture}
\begin{axis}[
    legend style={cells={align=left}},
    xbar stacked,
    width=0.8\columnwidth,
    height=3cm,
    xmin=0, xmax=100,
    bar width=1,
    xlabel={Proportion of Respondents (\%)},
    ytick={4,3,2,1},
    yticklabels={Prison ($n=99$),Mortgage ($n=97$),Job ($n=99$),Insurance ($n=97$)},
    xtick={0,20,40,60,80,100},
    tick label style={font=\footnotesize},
    label style={font=\footnotesize},
    ytick style={draw=none},
    axis x line*=bottom,
    axis y line*=left,
    ymin=0.6, ymax=4.4,
    nodes near coords,
    every node near coord/.append style={
    font=\footnotesize,
        color=black
    },
    nodes near coords align={center},
    legend style={
        overlay,
        font=\footnotesize,
        draw=none,
        at={(0.5,1.02)},
        xshift=-.5cm,  
        anchor=south,
        legend columns=2,
        cells={align=left},
        /tikz/every even column/.append style={column sep=0cm},
        row sep=.5pt
    }
]

\addplot+[xbar, fill=blue!40!cyan, draw=white]
coordinates {
    (21.2,4) [21.2]
    (33.0,3) [33.0]
    (12.1,2) [12.1]
    (34.0,1) [34.0]
};

\addplot+[xbar, fill=orange!75, draw=white]
coordinates {
    (15.2,4) [15.2]
    (27.8,3) [27.8]
    (11.1,2) [11.1]
    (20.6,1) [20.6]
};

\addplot+[xbar, fill=gray!25, draw=white]
coordinates {
    (15.2,4) [15.2]
    (9.3,3) [9.3]
    (13.1,2) [13.1]
    (24.7,1) [24.7]
};

\addplot+[xbar, fill=magenta!30, draw=white]
coordinates {
    (48.4,4) [48.4]
    (29.9,3) [29.9]
    (63.6,2) [63.6]
    (20.6,1) [20.6]
};

\legend{
\shortstack[l]{High bias; No chance of MI},
\shortstack[l]{Moderate bias; Slight chance of MI},
\shortstack[l]{Slight bias; Moderate chance of MI},
\shortstack[l]{No bias; High chance of MI}
}
\end{axis}
\end{tikzpicture}
}
\label{t2_fair_mi_revision}
\vspace{-.8cm}
\end{figure}

\addd{We provide the preference distribution in Figure~\ref{t2_fair_mi_revision}.}
\removee{
For the Prison and Job scenarios,  participants significantly preferred the fairness option. We provide the preference distribution in Figure~\ref{t2_fair_mi_revision}.} 
The fairness-maximizing option was selected by 63\% of participants in the Prison scenario; and within that 48\% were willing to accept a high risk of membership inference, with an additional 15\% accepting a moderate risk of membership inference. Similarly in the Job scenario, 76\%, preferred fairness-maximizing options. Breaking down the consequences, 63\% of selected a significant increase in membership inference risk and 13\% selected a moderate risk of membership inference. 
Conversely, participants prioritized privacy in the Mortgage and Insurance scenarios. In the Mortgage scenario, 60\% of participants prioritized privacy. The high bias was preferred by 33\% while 27\% accepted moderate bias in decision-making as their consequences. We observe a similar pattern in the Insurance scenario, where 54\% preferred privacy, and for consequences, 34\% accepted higher bias and 27\% preferred moderate bias.

\subsection{Fairness versus Distribution Inference}\label{result_t3}

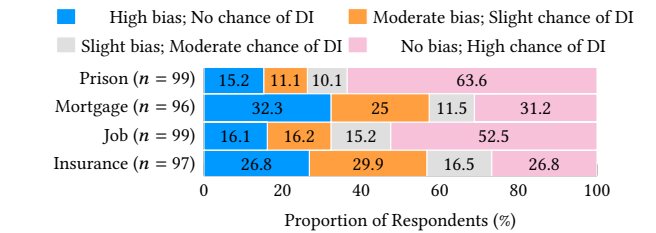
\begin{figure}[!htb]
\vspace{-.5cm}
\centering
\caption{Preferences of fairness versus distribution inference (DI) with sample size ($n<N$) for each scenario.}
\vspace{.8cm}
\begin{tikzpicture}
\begin{axis}[
    xbar stacked,
   width=0.8\columnwidth,
    height=3cm,
    xmin=0, xmax=100,
    bar width=1,
    xlabel={Proportion of Respondents (\%)},
    ytick={4,3,2,1},
    yticklabels={
        Prison ($n=99$),
        Mortgage ($n=96$),
        Job ($n=99$),
        Insurance ($n=97$)
    },
    xtick={0,20,40,60,80,100},
    tick label style={font=\footnotesize},
    label style={font=\footnotesize},
    ytick style={draw=none},
    axis x line*=bottom,
    axis y line*=left,
    ymin=0.6,
    ymax=4.4,
    nodes near coords,
    every node near coord/.append style={
        font=\footnotesize,
        color=black
    },
    nodes near coords align={center},
  legend style={
        overlay,
        font=\footnotesize,
        draw=none,
        at={(0.5,1.02)},
        xshift=-.7cm,  
        anchor=south,
        legend columns=2,
        cells={align=left},
        /tikz/every even column/.append style={column sep=0cm},
        row sep=.5pt
    }
]
\addplot+[xbar, fill=blue!40!cyan, draw=white]
coordinates {
    (15.2,4) [15.2]
    (32.3,3) [32.3]
    (16.1,2) [16.1]
    (26.8,1) [26.8]
};

\addplot+[xbar, fill=orange!75, draw=white]
coordinates {
    (11.1,4) [11.1]
    (25,3) [25]
    (16.2,2) [16.2]
    (29.9,1) [29.9]
};

\addplot+[xbar, fill=gray!25, draw=white]
coordinates {
    (10.1,4) [10.1]
    (11.5,3) [11.5]
    (15.2,2) [15.2]
    (16.5,1) [16.5]
};

\addplot+[xbar, fill=magenta!30, draw=white]
coordinates {
    (63.6,4) [63.6]
    (31.2,3) [31.2]
    (52.5,2) [52.5]
    (26.8,1) [26.8]
};

\legend{\shortstack[l]{High bias; No chance of DI},\shortstack[l]{Moderate bias; Slight chance of DI},\shortstack[l]{Slight bias; Moderate chance of DI},\shortstack[l]{No bias; High chance of DI}}

\end{axis}
\end{tikzpicture}
\label{t3_fair_di_revision}
\vspace{-.5cm}
\end{figure}

\removee{Participants showed a significant preference for choosing fairness in Prison sentencing and Job hiring, and their preference distribution is shown in Figure~\ref{t3_fair_di_revision}.}
 In the Prison scenario, 73\% selected fairness-maximizing options, with 63\% accepting a high chance of distribution inference and 10\% selecting the moderate risk of distribution inference, \addd{as shown in Figure~\ref{t3_fair_di_revision}}. Similarly, in the Job scenario, 67\% of respondents prefer fairness-prioritizing options, where 52\% preferred the high risk of distribution inference and 15\% preferred the moderate risk. 
In the Mortgage approval scenario, 57\% of participants chose privacy-preferred options, broken down as 32\% that accepted a strong bias towards one group, and 25\% accepted a moderate bias. In the Insurance scenario, 56\% selected privacy-maximizing options, 26\% accepted high bias, and 29\% accepted moderate bias.

\subsection{Fairness versus Security}\label{result_t4}
\removee{In the Prison and Mortgage scenarios, participants showed a significant preference for choosing the highest level of security. Their preference distribution is shown in Figure \ref{t4_fair_rob_revision}.}
\addd{Figure \ref{t4_fair_rob_revision} shows the participants' preference distribution.}
In the Prison scenario, 60\% of participants preferred security. Broken down by consequences, 46\% accepted high bias and 14\% accepted moderate bias as the consequences. Security options were similarly preferred by 67\% in the Mortgage scenario and high bias being accepted by 42\% and 25\% accepting moderate bias. In the Insurance scenario, 55\% selected the option prioritizing security, with 36\% accepting high bias and 19\% accepting moderate bias. Unlike the other three scenarios, in the Job scenario 51\% of participants preferred fairness-maximizing options with their consequences being 32\% preferred high manipulation risk and 19\% preferred moderate manipulation risk.

\begin{figure}[H]
\centering
\vspace{-.5cm}
\caption{Preferences of fairness versus manipulation risk (MR) with sample size ($n<N$) for each scenario.}
\vspace{.8cm}
\label{t4_fair_rob_revision}
\begin{tikzpicture}
\begin{axis}[
    xbar stacked,
   width=0.8\columnwidth,
    height=3cm,
    xmin=0, xmax=100,
    bar width=1,
    xlabel={Proportion of Respondents (\%)},
    ytick={4,3,2,1},
    yticklabels={
        Prison ($n=100$),
        Mortgage ($n=96$),
        Job ($n=99$),
        Insurance ($n=94$)
    },
    xtick={0,20,40,60,80,100},
    tick label style={font=\footnotesize},
    label style={font=\footnotesize},
    ytick style={draw=none},
    axis x line*=bottom,
    axis y line*=left,
    ymin=0.6,
    ymax=4.4,
    nodes near coords,
    every node near coord/.append style={
        font=\footnotesize,
        color=black
    },
    nodes near coords align={center},
      legend style={
        overlay,
        font=\footnotesize,
        draw=none,
        at={(0.5,1.02)},
        xshift=-.5cm,  
        anchor=south,
        legend columns=2,
        cells={align=left},
        /tikz/every even column/.append style={column sep=0cm},
        row sep=.5pt
    }
]
\addplot+[xbar, fill=blue!40!cyan, draw=white]
coordinates {
    (46,4) [46]
    (42.7,3) [42.7]
    (29.3,2) [29.3]
    (36.2,1) [36.2]
};

\addplot+[xbar, fill=orange!75, draw=white]
coordinates {
    (14,4) [14]
    (25,3) [25]
    (19.2,2) [19.2]
    (19.1,1) [19.1]
};

\addplot+[xbar, fill=gray!25, draw=white]
coordinates {
    (22,4) [22]
    (16.7,3) [16.7]
    (19.2,2) [19.2]
    (22.3,1) [22.3]
};

\addplot+[xbar, fill=magenta!30, draw=white]
coordinates {
    (18,4) [18]
    (15.6,3) [15.6]
    (32.3,2) [32.3]
    (22.3,1) [22.3]
};

\legend{\shortstack[l]{High bias; No MR},\shortstack[l]{Moderate bias; Slight MR},\shortstack[l]{Slight bias; Moderate MR},\shortstack[l]{No Bias; High MR}}

\end{axis}
\end{tikzpicture}
\vspace{-.5cm}
\end{figure}
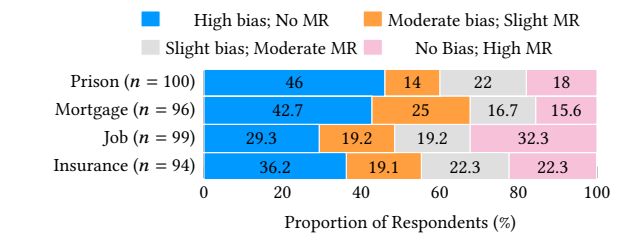
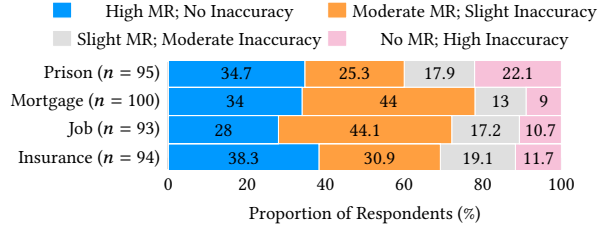
\begin{figure}[!htb]
\centering
\vspace{-.5cm}
\caption{Preferences for manipulation risk versus accuracy with sample size ($n<N$) for each scenario.}
\vspace{.8cm}
\begin{tikzpicture}
\begin{axis}[
    legend style={cells={align=left}},
    xbar stacked,
  width=0.8\columnwidth,
    height=3cm,
    xmin=0, xmax=100,
    bar width=1,
    xlabel={Proportion of Respondents (\%)},
    ytick={4,3,2,1},
    yticklabels={Prison ($n=95$),Mortgage ($n=100$),Job ($n=93$),Insurance ($n=94$)},
    xtick={0,20,40,60,80,100},
    tick label style={font=\footnotesize},
    label style={font=\footnotesize},
    ytick style={draw=none},
    axis x line*=bottom,
    axis y line*=left,
    ymin=0.6, ymax=4.4,
    nodes near coords,
    every node near coord/.append style={
    font=\footnotesize,
        color=black
    },
    nodes near coords align={center},
      legend style={
        overlay,
        font=\footnotesize,
        draw=none,
        at={(0.5,1.02)},
        xshift=-.5cm,  
        anchor=south,
        legend columns=2,
        cells={align=left},
        /tikz/every even column/.append style={column sep=0cm},
        row sep=.5pt
    }
]
\addplot+[xbar, fill=blue!40!cyan, draw=white]
coordinates {
    (34.7,4) [34.7]
    (34,3) [34]
    (28,2) [28]
    (38.3,1) [38.3]
};

\addplot+[xbar, fill=orange!75, draw=white]
coordinates {
    (25.3,4) [25.3]
    (44,3) [44]
    (44.1,2) [44.1]
    (30.9,1) [30.9]
};

\addplot+[xbar, fill=gray!25, draw=white]
coordinates {
    (17.9,4) [17.9]
    (13,3) [13]
    (17.2,2) [17.2]
    (19.1,1) [19.1]
};

\addplot+[xbar, fill=magenta!30, draw=white]
coordinates {
    (22.1,4) [22.1]
    (9,3) [9]
    (10.7,2) [10.7]
    (11.7,1) [11.7]
};

\legend{\shortstack[l]{High MR; No Inaccuracy},\shortstack[l]{Moderate MR; Slight Inaccuracy},\shortstack[l]{Slight MR; Moderate Inaccuracy},\shortstack[l]{No MR; High Inaccuracy}}

\end{axis}
\end{tikzpicture}
\label{t5_rob_accu_revision}
\vspace{-.5cm}
\end{figure}

\subsection{Security versus Accuracy}
\remove{Across all four scenarios, a larger proportion of participants selected accuracy-maximizing options over security.} \removee{In all scenarios except Prison, participants significantly chose the accuracy-maximizing option. Their preference distribution is shown in Figure~\ref{t5_rob_accu_revision}.} 
In the Prison scenario, 60\% of participants chose accuracy-maximizing options, with 34\% accepting the risk of high manipulation and 25\% accepting a moderate manipulation risk, \addd{shown in Figure~\ref{t5_rob_accu_revision}}. A similar pattern appeared in the Mortgage scenario, where 74\% preferred accuracy-maximizing options, with 34\% accepting high manipulation risk and 40\% accepting moderate manipulation risk. 
The Job scenario showed a comparable trend, with 72\% opting for accuracy-maximizing choices, which broken down by consequences is 28\% tolerating high manipulation risk, and 44.1\% moderate manipulation risk. In the Insurance scenario, 68\% choose accuracy-maximizing options, which broken down is 38.3\% accepted high manipulation risk, and 30\% accepted moderate manipulation risk. Altogether, these results indicate a broad willingness among participants across scenarios to accept the risk of getting an inaccurate prediction if it prevented manipulation.

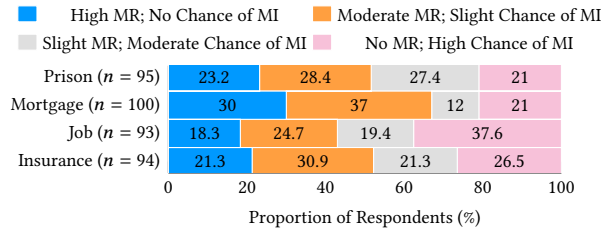
\begin{figure}[H]
\centering
\vspace{-.5cm}
\caption{Preferences for manipulation risk (MR) versus membership inference (MI) with sample size ($n<N$).}
\vspace{.8cm}
\begin{tikzpicture}
\begin{axis}[
    legend style={cells={align=left}},
    xbar stacked,
  width=0.8\columnwidth,
    height=3cm,
    xmin=0, xmax=100,
    bar width=1,
    xlabel={Proportion of Respondents (\%)},
    ytick={4,3,2,1},
    yticklabels={Prison ($n=95$),Mortgage ($n=100$),Job ($n=93$),Insurance ($n=94$)},
    xtick={0,20,40,60,80,100},
    tick label style={font=\footnotesize},
    label style={font=\footnotesize},
    ytick style={draw=none},
    axis x line*=bottom,
    axis y line*=left,
    ymin=0.6, ymax=4.4,
    nodes near coords,
    every node near coord/.append style={
    font=\footnotesize,
        color=black
    },
    nodes near coords align={center},
    legend style={
        overlay,
        font=\footnotesize,
        draw=none,
        at={(0.5,1.02)},
        xshift=-.7cm,  
        anchor=south,
        legend columns=2,
        cells={align=left},
        /tikz/every even column/.append style={column sep=0cm},
        row sep=.5pt
    }
]
\addplot+[xbar, fill=blue!40!cyan, draw=white]
coordinates {
    (23.2,4) [23.2]
    (30,3) [30]
    (18.3,2) [18.3]
    (21.3,1) [21.3]
};

\addplot+[xbar, fill=orange!75, draw=white]
coordinates {
    (28.4,4) [28.4]
    (37,3) [37]
    (24.7,2) [24.7]
    (30.9,1) [30.9]
};

\addplot+[xbar, fill=gray!25, draw=white]
coordinates {
    (27.4,4) [27.4]
    (12,3) [12]
    (19.4,2) [19.4]
    (21.3,1) [21.3]
};

\addplot+[xbar, fill=magenta!30, draw=white]
coordinates {
    (21,4) [21]
    (21,3) [37.6]
    (37.6,2) [19.4]
    (26.5,1) [26.5]
};

\legend{{High MR; No Chance of MI}, {Moderate MR; Slight Chance of MI},{Slight MR; Moderate Chance of MI},{No MR; High Chance of MI}}

\end{axis}
\end{tikzpicture}
\label{t6_rob_mi_revision}
\vspace{-.5cm}
\end{figure}

\subsection{Security versus Membership Inference}
\remove{Participants show varying intensities in their preferences for manipulating risk and privacy protection of membership inference.} 
\removee{ 
Scenario preference distributions are shown in Figure~\ref{t6_rob_mi_revision}, and}

\removee{ participants did not choose any option significantly more than others.} 
\addd{Figure~\ref{t6_rob_mi_revision} shows the distribution.} In Mortgage, 67\% chose privacy-maximizing options, with 30\% willing to take a high risk and 37\% willing to accept a moderate risk of manipulation. In the Prison and Insurance scenarios, preferences are split between privacy and security. In particular, the Prison scenario shows 51\% preferring privacy-maximizing options, with 23\% accepting a high risk and 28\% accepting a moderate risk of manipulation.  In the Insurance scenario, 52\% selected privacy-maximizing options, with 21\% choosing a high likelihood of manipulation risk and 30\% selecting a moderate likelihood of manipulation risk. The Job scenario had 57\% select the security-maximizing option, where 37\% selected high chance of MI, and 19\% selected a moderate chance of MI.

\subsection{Security versus Distribution Inference}
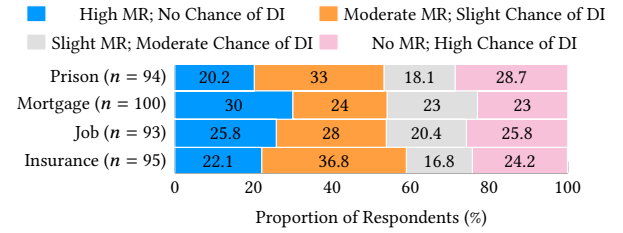
\begin{figure}[H]
\vspace{-.5cm}
\centering
\caption{Preferences for manipulation risk (MR) versus distribution inference (DI) with sample size ($n<N$).}
\vspace{.8cm}
\begin{tikzpicture}
\begin{axis}[
    legend style={cells={align=left}},
    xbar stacked,
  width=0.8\columnwidth,
    height=3cm,
    xmin=0, xmax=100,
    bar width=1,
    xlabel={Proportion of Respondents (\%)},
    ytick={4,3,2,1},
    yticklabels={Prison ($n=94$),Mortgage ($n=100$),Job ($n=93$),Insurance ($n=95$)},
    xtick={0,20,40,60,80,100},
    tick label style={font=\footnotesize},
    label style={font=\footnotesize},
    ytick style={draw=none},
    axis x line*=bottom,
    axis y line*=left,
    ymin=0.6, ymax=4.4,
    nodes near coords,
    every node near coord/.append style={
    font=\footnotesize,
        color=black
    },
    nodes near coords align={center},
      legend style={
            overlay,
            font=\footnotesize,
            draw=none,
            at={(0.5,1.02)},
            xshift=-.7cm,  
            anchor=south,
            legend columns=2,
            cells={align=left},
            /tikz/every even column/.append style={column sep=0cm},
            row sep=.5pt
        }
]
\addplot+[xbar, fill=blue!40!cyan, draw=white]
coordinates {
    (20.2,4) [20.2]
    (30,3) [30]
    (25.8,2) [25.8]
    (22.1,1) [22.1]
};

\addplot+[xbar, fill=orange!75, draw=white]
coordinates {
    (33,4) [33]
    (24,3) [24]
    (28,2) [28]
    (36.8,1) [36.8]
};

\addplot+[xbar, fill=gray!25, draw=white]
coordinates {
    (18.1,4) [18.1]
    (23,3) [23]
    (20.4,2) [20.4]
    (16.8,1) [16.8]
};

\addplot+[xbar, fill=magenta!30, draw=white]
coordinates {
    (28.7,4) [28.7]
    (23,3) [23]
    (25.8,2) [25.8]
    (24.2,1) [24.2]
};

\legend{\shortstack[l]{High MR; No Chance of DI},\shortstack[l]{Moderate MR; Slight Chance of DI},\shortstack[l]{Slight MR; Moderate Chance of DI},\shortstack[l]{No MR; High Chance of DI}}

\end{axis}
\end{tikzpicture}
\label{t7_rob_di_revision}
\vspace{-.5cm}
\end{figure}

\remove{The Prison, Mortgage, and Job scenario responses were relatively evenly distributed between privacy and security with a higher proportion selecting privacy-maximizing options,} \removee{We did not find
any significant difference in preference, and their preference distribution is shown in Figure \ref{t7_rob_di_revision}} In the Prison scenario, 53\% selected privacy-preferred options, 20\% accepted a high risk, and 33\% accepted a moderate risk of adversarial manipulation, \addd{as shown in Figure \ref{t7_rob_di_revision}}. A similar distribution is observed in the Mortgage scenario, where 54\% selected privacy-maximizing options, 30\% accepted high risk, and 24\% accepted moderate risk of manipulation. The Job scenario shows a comparable pattern, with 53\% selecting privacy-preferred options, 25\% accepting high risk, and 28\% preferring moderate risk of manipulation.
In the Insurance scenario, 58.9\% selected privacy-maximizing options,  22\% selected high-risk, and 36\% selected moderate-risk of manipulation.

\subsection{Security versus Data Reconstruction}  
\begin{figure}[H]
\vspace{-.5cm}
\centering
\caption{Preferences for manipulation risk (MR) versus data reconstruction (DR) risk with sample size ($n<N$).}
\vspace{.8cm}
\begin{tikzpicture}
\begin{axis}[
    legend style={cells={align=left}},
    xbar stacked,
  width=0.8\columnwidth,
    height=3cm,
    xmin=0, xmax=100,
    bar width=1,
    xlabel={Proportion of Respondents (\%)},
    ytick={4,3,2,1},
    yticklabels={Prison ($n=93$),Mortgage ($n=100$),Job ($n=92$),Insurance ($n=94$)},
    xtick={0,20,40,60,80,100},
    tick label style={font=\footnotesize},
    label style={font=\footnotesize},
    ytick style={draw=none},
    axis x line*=bottom,
    axis y line*=left,
    ymin=0.6, ymax=4.4,
    nodes near coords,
    every node near coord/.append style={
    font=\footnotesize,
        color=black
    },
    nodes near coords align={center},
      legend style={
        overlay,
        font=\footnotesize,
        draw=none,
        at={(0.5,1.02)},
        xshift=-.7cm,  
        anchor=south,
        legend columns=2,
        cells={align=left},
        /tikz/every even column/.append style={column sep=0cm},
        row sep=.5pt
    }
]
\addplot+[xbar, fill=blue!40!cyan, draw=white]
coordinates {
    (21.5,4) [21.5]
    (28,3) [28]
    (34.8,2) [34.8]
    (26.6,1) [26.6]
};

\addplot+[xbar, fill=orange!75, draw=white]
coordinates {
    (28,4) [28]
    (34,3) [34]
    (23.9,2) [23.9]
    (28.7,1) [28.7]
};

\addplot+[xbar, fill=gray!25, draw=white]
coordinates {
    (16.1,4) [16.1]
    (9,3) [9]
    (20.7,2) [20.7]
    (14.9,1) [14.9]
};

\addplot+[xbar, fill=magenta!30, draw=white]
coordinates {
    (34.4,4) [34.4]
    (29,3) [29]
    (20.6,2) [20.6]
    (29.8,1) [29.8]
};

\legend{\shortstack[l]{High MR; No Chance of DR},\shortstack[l]{Moderate MR; Slight Chance of DR},\shortstack[l]{Slight MR; Moderate Chance of DR},\shortstack[l]{No MR; High Chance of DR}}

\end{axis}
\end{tikzpicture}
\label{t8_rob_recon_revision}
\vspace{-.5cm}
\end{figure}
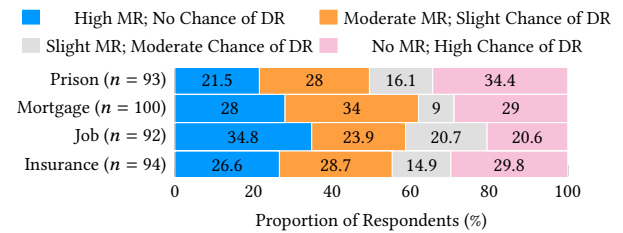

\remove{A greater proportion of participants chose protection from data reconstruction over protection against adversarial manipulation.} \removee{We did not find any scenario where participants significantly preferred one option over others.} Preference distribution is shown in Figure \ref{t8_rob_recon_revision}.
In the Mortgage scenario, 62\% selected privacy-maximizing options, and correspondingly 28\% accepted the high risk of manipulation and 34\% accepted a moderate risk. The Job scenario had 58.7\% preferring privacy, where 34\% selected the high risk of manipulation and 23\% selecting moderate risk. For the Insurance scenario, 55\% chose privacy-maximizing options, where 26\% accepted a high manipulation risk and 28\% accepting a moderate risk. In the Job scenario, preferences are divided as 49\% selected options with the lowest privacy risk where 21\% accepted a high risk of manipulation, and 16\% accepted a moderate risk of manipulation.


\section{Tradeoffs Preference Across Scenarios} \label{tradeoff_across_scenario}
\addd{To determine whether participants' preference distributions vary across scenarios, we ran the chi-square test for each tradeoff across the four scenarios. We report the tradeoffs where participants show significant differences and, for the pairs that differ (details in Appendix~\ref{appendix_tradeoffs_quant}). In the tradeoff between fairness and MI, participants showed significant differences across scenarios, with $ p <.001$ and $\chi^2=.377$. Specifically, the following pairs showed a significant difference: (Insurance, Job), (Insurance, Prison), (Job, Mortgage) and (Mortgage, Prison). We also found significant differences in fairness versus DI, with $p < .001$ and $\chi^2=41.843$. The following pairs showed a significant difference: (Insurance, Job), (Insurance, Prison), (Job, Mortgage), and (Mortgage, Prison). In all other tradeoffs, we did not find any significant difference in preference across the scenarios.}

\section{Findings for Tradeoff Considerations}\label{sect:considerationstradeoffs}
We use our qualitative analysis to elucidate participants' considerations when prioritizing an objective in a tradeoff. \add{Participants' priorities among the objectives were shaped by the specific ADM context presented in each scenario.}
We organize our findings, the themes, by the values, concerns, and reasoning provided by participants in their freeform responses. 

\subsection{Concerns Beyond Technical Fairness Metrics}
Participants' perceptions of fairness go beyond ensuring similar outcomes across demographic groups, incorporating privacy, accuracy, and security risks. Fairness for the participants includes proportionality, where outcomes are expected to align with an individual’s risk or qualifications, with deviations perceived as unfair.

\paragraph{Lack of privacy protection leads to discrimination} \label{mi_discriminatory}
Participants perceived that disclosing specific attributes can negatively influence decision-making, particularly when it triggers the correlation of such information with historical biases.
Whether the disclosed information is directly related to the participants (e.g., membership inference) or more broadly about the training dataset (e.g., distribution inference), such disclosures were seen as a potential discriminatory factor that is ``\ldots, undermining the fairness that AI was supposed to promote" [P191, Mortgage].
For example, in the Mortgage context, participants expressed that ``it is better to keep client demographic private so they have a fair chance at obtaining the loan'' [P505, Mortgage].
Participants noted that, even though demographic statistics are not directly related to an individual user, they can ``\ldots lead to indirect forms of discrimination or reinforcing [demographic] stereotypes" [P403, Mortgage].

\paragraph{Susceptibility to manipulation is perceived as a source of discrimination}\label{robustness_discrimination}
Participants emphasized that 
a lack of security introduces opportunities for manipulation, where ``\ldots [human] could intentionally penalize groups, leading to targeted discrimination \ldots'' [P185, Prison]. As a result, participants highlighted that ``\ldots preventing manipulation is crucial to maintain fairness \ldots'' [302, Mortgage] and preserve ``system integrity'' [P200, Prison]. Participants further noted that when systems are easily manipulated, they can be exploited for organizational gain, as, ``if the AI is too easy to manipulate, employees will exploit it to give customers higher premiums for greater profit. This is unfair and possibly illegal'' [P685, Insurance]. Overall, participants conceptualize security as more than a technical property of ADM systems, viewing it as a protection against discriminatory outcomes that arise from human manipulation of ADM systems.

\paragraph{Inaccuracy results in discrimination}
Participants closely linked perceptions of fairness to the accuracy of ADM outputs, noting that ``\ldots if people are getting [results] that are in any way incorrect, that is unfair all around'' [P99, Insurance]. This concern appeared in all decision-making contexts, with one example being: ``\ldots low-risk clients might pay too much \ldots which is not fair to anyone'' [P323, Mortgage]. In all of these contexts, inaccurate outcomes were perceived as imposing unjust burdens on individuals.

Overall, participants did not treat fairness as an isolated construct; rather, they framed accuracy, security against adversarial manipulation, and privacy as intertwined with fairness. In particular, inaccuracy, lack of security, and membership inference were perceived as direct sources of unfairness when they led to tangible harms, while distribution inference was seen as potentially triggering unconscious bias or reinforcing demographic-based stereotypes.

\subsection{Contextual Influences on Prioritization}
Depending on the sensitive information in the decision-making scenario, participants perceive some information as contextually appropriate to reveal, even through MI, if it aids in decision-making.  

\paragraph{Participants believe that specific Membership inference helps make informed decisions in certain scenarios}
Some participants perceived that membership inference is acceptable when it can lead to informed decision-making. For example, in the Job scenario, criminal history is perceived as a crucial piece of information. 
Alongside the information being crucial, such participants expressed concern that if the ADM system failed to reveal criminal history then ``\ldots the [ADM] would hire those with a criminal background, which could be bad for the work force of the company'' [P564, Job]. 


\paragraph{Participants consider accepting the risk of revealing certain membership inference to ensure fairness and security} 
An influence on participants' prioritization was whether information disclosure was perceived as inevitable in a given decision-making context. 
For example, in the Prison scenario, participants noted that the justice system gathers a certain amount of information about the prisoner; thus, ``wealth is easy to objectively measure in this instance'' [P602, Prison]. Despite acknowledging the availability of certain attributes, participants still articulated normative boundaries, stating that they ``do not think an individual's wealth should matter when delivering justice \ldots'' [P69, Prison]. 
Similarly, 
in the Job scenario participants perceived that ``criminal backgrounds should be taken into consideration during the hiring process \ldots'' [P136, Job] and thus it is fine to be revealed as disclosures of criminal history are a regular practice in hiring. 
Overall, participants expressed a hierarchy of harms, in which violations of fairness and security were considered more critical than certain privacy risks, particularly when information disclosure was deemed contextually appropriate or unavoidable.
\paragraph{Participants consider accepting the risk of revealing certain distributive inference to ensure fairness and security} Participants perceived distributive inference as involving non-sensitive, aggregated information, and, in some cases, to even be a form of transparency.
For example, participants framed demographic information as useful for monitoring fairness, suggesting that in the hiring context ``knowing the number of women in the dataset could help the [company] monitor diversity and ensure fair hiring practices \ldots '' [P756, Job]. 
While there was some acknowledgment of the benefits of protecting aggregate demographic data, participants emphasized that preventing adversarial manipulation and ensuring fair outcomes were higher-order priorities. 
 In short, a prevailing view was that ``revealing demographic proportions like gender in training data might not directly harm individual applicants, but allowing easy AI manipulation would introduce greater harm through unfair decisions and biased interest rates'' [P370, Mortgage]. Overall, participants prioritized security and fairness over protecting aggregate demographic information, particularly when such information was perceived as non-sensitive and contextually appropriate to disclose.

\subsection{Accepting Bias for the Greater Good}
Participants wanted to prevent harm and system manipulation while benefiting stakeholders
However, conditionally, participants were willing to accept some bias despite it also being a harm.
\paragraph{Participants show acceptance of bias if it does not imply group marginalization}
Participants tended to accept bias where it was in a form of favoring a group, with the perspective that ``\ldots favoring one group is better than marginalizing the group. I would rather it be lenient than harsh \ldots'' [P750, Prison]. Favoring one group was perceived as causing minimal harm, as it primarily benefits the favored group without necessarily, directly, disadvantaging others. 
While our participants perceived favoritism as not as harmful as more direct discriminatory behavior, we acknowledge that the consequences are much the same. 

\paragraph{Participants consider favoritism towards one group acceptable when it leads to benefiting the majority of stakeholders}
In the accuracy versus fairness and security respectively, participants prioritized minimizing harm to the larger population. Across the scenarios, they observed that the group favored by the system or affected by the risk of adversarial manipulation accounted for only a small portion of the population. Thereby, they prioritized accuracy, since ``\ldots at least the algorithm provides correct [result] to everyone else \ldots'' [P145, Insurance].
Participants generally viewed fairness and security as important, but not at the expense of accurate decisions. For example, 
they acknowledge that while ``\ldots eliminating the bias is important, doing so at the cost of widespread inaccuracies \ldots would be a greater injustice'' [P245, Prison]. Similarly, in the security versus accuracy tradeoffs, when we sacrifice accuracy to achieve security, participants expressed, ``\ldots it is a zero-sum game. It does not matter if the [ADM] is impenetrable from manipulation if the [results] are highly inaccurate, because inaccuracy is what the manipulation protection is trying to protect" [P402, Mortgage].

\paragraph{Participants consider accepting bias due to training data to stop adversarial manipulation}
In the fairness-versus-security trade-off, our findings show that participants perceived adversarial manipulation as deliberate discrimination and found it unacceptable.
Participants considered it``\ldots more acceptable for [ADMS] to have a reasonable amount of bias than to have the system susceptible to deliberate misuse \ldots'' [P223, Insurance]. Bias due to training data was considered a technical flaw beyond control, whereas human manipulation was perceived as a controllable problem and thus less acceptable. 

\subsection{Consequences Beyond Biased Outcomes}
Participants expressed concern for consequences beyond the stated bias. They brought up that there were also implicit harms, including inaccuracy and an undermining of the systems validity. 

\paragraph{Participants consider unfair outcomes as ADM's inaccuracy}
Recall from Section~\ref{result_t1}, participants’ view that inaccurate outputs correspond to unfair outcomes. Similarly, participants framed fairness and accuracy as inherently interconnected, stating, ``if someone or something is showing any type of bias, then there is a huge issue with fairness and accuracy \ldots'' [P84, Prison]. This perspective implies that a system that embeds structural bias is not only unfair, but seen as producing unreliable and incorrect decisions.

\paragraph{Participants perceived the risk of manipulation as undermining the system's integrity}
Participants extended their concerns about the risk of manipulation beyond unfair decision outcomes to highlight broader system-level vulnerabilities. In particular, participants expressed that ``preventing manipulation by employees is critical to maintaining system integrity \ldots'' [P198, Prison], and ``if the [ADM] is vulnerable to employee tampering, it undermines the entire system's \ldots security \ldots'' [P185, Mortgage]. 

\paragraph{Participants recognize the potential risk that can come from exposing distribution inference}
Participants emphasized that seemingly non-sensitive aggregate demographic statistics still introduce risks, including ``\ldots ethical issues about exposing group-level data without consent" [P367, Mortgage]. 
Disclosures from distribution inference were noted by participants to facilitate secondary uses, 
``\ldots which would undermine efforts to ensure fairness and privacy in financial systems" [P167, Mortgage].
They further observed 
``\ldots aggregated data can be misused to reinforce exclusionary practices \ldots'' [P185, Mortgage], or enable ``gender profiling or unintended use of this information to adjust pricing strategies \ldots'' [P221, Insurance].

Overall, participants highlighted that both manipulation risks and aggregate data disclosures can introduce system-level issues resulting in unintended risks beyond individual decision outcomes.

\subsection{Negative Impact on Stakeholders Matters}
In all cases, when participants prioritize one objective over another, they consider its affect on end users, organizations, and society.

\paragraph{Loss of trust}
Inaccuracy, bias, and the ADM system's susceptibility to manipulation lead to loss of trust in ADM systems. Inaccurate decisions may lead customers to disengage, as, ``people will be upset about the incorrect interest rates'' [P476, Mortgage].
Unequal treatment was also perceived as undermining public trust, particularly when such discrimination emerged from an institution whose primary goal is to ensure justice. Participants expressed that ``\ldots unfair sentencing based on looks \ldots could negatively impact the trust in the justice system'' [P202, Prison]. 
In hiring, when unsuitable candidates are hired, it affects the individuals who deserved the job most, as well as workforce quality and the company's reputation by ``undermining trust in the hiring process'' [P336, Job]. Finally, any privacy violation, either revealing membership inference, distribution inference, or an adversarial party recovering the training data, is perceived as harming trust, which
``\ldots would compromise the entire system and generate a lot of negative press about the company and the AI \ldots" [P747, Insurance].


\paragraph{Reinforcing historical injustice}
The consequences of membership inference were not limited to the inference itself. Rather it was perceived that membership inference ``\ldots compromises client privacy and can deter minority applicants from applying at all" [P167, Mortgage].
Essentially, the concern was that the group of people who are historically disadvantaged by the system will be further disadvantaged in an automated way.
In the hiring context, if the company focused less on privacy and allowed easy identification of applicants with criminal records, ``\ldots it could lead to \ldots discouraging qualified individuals from applying and harming diversity and inclusion efforts" [P756, Job].


\paragraph{Financial loss}
The prioritization of different objectives can cause financial damage to both users and the institution behind the ADM system.
Across scenarios, participants framed inaccuracy as disproportionately harming those who should otherwise benefit from fair outcomes. In the Mortgage scenario this was expressed as``higher interest rates for undeserving people'' [P425, Mortgage]. 
Incorrect risk assessments also directly affect the profitability of the institution running the ADM, as a high-risk person receiving ``a lower interest rate will be costly for the company \ldots [and] the company will not be able to profit well'' [P337, Mortgage]. In hiring, ``if [the company] \ldots hire someone wrong \ldots this is expensive and time consuming'' [P535, Job]. 
Finally, the loss of data privacy could indirectly lead to financial harms as it ``\ldots could give competing companies extra inside information" [P717, Insurance]. 

\paragraph{Legal liability}
Beyond financial impacts, participants raised concerns that inaccurate outcomes may expose institutions using ADM systems to legal consequences.
Participants indicated that favoring one group is``\ldots violating anti-discrimination laws and damaging the credibility of both the [ADM] system and the [institution].'' [P671, Prison]. 
There are also legal concerns regarding privacy failures,
where it was stated that, ``\ldots it is a HIPAA violation for the [ADM] to reveal [health] issues \ldots'' [P249, Insurance]. 


\paragraph{Societal impact}
The Prison scenario in particular raised concerns from participants about the societal impact of ADM's incorrect decisions. Inaccuracy in the Prison scenario was described as life-altering, where it ``would unnecessarily make an innocent person spend a longer time in prison'' [P172, Prison].
Inaccuracies were seen as harmful to both individuals and society, and
``if the AI were easy to manipulate, violent or dangerous criminals could be released back into the public'' [P159, Prison].

\section{Discussion}

The following is our discussion of actions for concerns about privacy, fairness, and security in the design and governance of ADM systems; to increase awareness and mitigation of associated risks.

\paragraph{Folk tales of effectiveness and efficiency}
Participants' perceptions of ADM focused on the benefits of using a machine,
as well as showing a limited and sometimes inaccurate mental model of how ADM works.
For example, participants' expressed that ADM systems handle large-scale data without error and make fact-driven decisions better than humans when the reality is more nuanced. 
This aligns with previous findings that non-experts use simplified, sometimes misleading mental models to describe what algorithms do and how they work, often without an accurate understanding of system limitations~\cite{ytre2021folk, ridley2024informing, eslami2017careful}. 
We can view these behaviors through the notion of ``folk theories'', which suggest that when people have limited knowledge of ADM, they are more likely to exhibit over-reliance on and have excessive confidence in the outcomes the of ADM system~\cite{horowitz2024bending}.
Consequently, while policymakers advocate for the inclusion of human oversight and expectations in the design of ADM system, it can be more challenging and often risky if the individuals
possess only a limited understanding of what they are assessing. Addressing this requires accurate reporting of system limitations from ADM providers alongside better ADM literacy among both policymakers and the populace.



\paragraph{Sensitivity of information is enmeshed with potential harm}
Our participants' reasoning in regards to sensitive information and harm is reflective of principles of privacy calculus models~\cite{culnan1999informationPrivacyCalculus} as well as boundary regulation theory~\cite{altman1975environment} and Nissenbaum’s contextual integrity framework~\cite{nissenbaum2004privacy}. Participants conceptualized information sensitivity primarily in relation to its potential to enable discrimination and how the information is relevant to an individual. Participants were more willing to accept privacy risks when the information was perceived as less sensitive, normatively appropriate, or socially acceptable within the given context. 


From a design and governance perspective, these findings suggest the need for more context-sensitive data practices in the deployment of ADM systems. One practical implication is the use of structured dataset documentation, such as datasheets for the dataset~\cite{gebru2021datasheets}, that explicitly account for stakeholder perceptions and contextual expectations. Such documentation should not only describe what data are used, but also show how specific attributes may function as proxies for discrimination, how they align with societal norms in the application domain, and how users perceive their appropriateness. In high-stakes settings, where enhancing privacy may come at the cost of other system objectives, these trade-offs should be made explicit and grounded in both social acceptability and perceived cost–benefit considerations. 


\paragraph{Deceptive linguistics is a concern for influencing participants' preferences}
Our participants expressed that favoring one group is not as bad as marginalizing another, even though both are unfair outcomes. 
This more positive view of favoritism aligns with the notion that linguistic choices shape how participants reason about their moral judgments and risk assessments, even when the outcomes are objectively equivalent~\cite{tversky1981framing}. 
For example, recall the Amazon hiring case where the algorithm favored men
~\cite{reutersAmazonBias2018}. Women were unfairly excluded, because of the favoritism.
Thus, this is a potential deceptive communication pattern, where the framing of tradeoffs, as favoring rather than marginalizing, can influence participants' decisions, despite the similarity in terms of exclusionary outcomes.

\paragraph{Presented bias in comparison to existing societal bias}
\add{By using more neutral proxy attributes, such as hair color, we avoid invoking existing societal stereotypes or counter-stereotypes about who is or is not favored, minimizing confounding effects associated with protected characteristics.
This choice avoids having historically social power holders be favored over historically marginalized groups and having historically marginalized groups favored over historically social power holders, which could have shifted participants' perspectives to focus on whether the presented setting aligned with their experience of the world rather than focusing on considering our settings. This indirect and hypothetical framing created distance between participants and the sensitive issue, allowing them to first perceive the underlying principles of algorithmic discrimination without requiring them to take an explicit position on a socially charged category. Given that even our more neutral proxy attributes indicate that participants contextually prioritize among tradeoffs, the importance of context should be accounted for in all design and regulatory decisions regarding ADMs.}

\paragraph{On the limitations of abstraction of human values}

Participants valued humans' subjective understanding and the ability to interpret circumstances and social context, qualities they viewed as fundamental to fair and legitimate decision-making.
While this supports human oversight as a component of trustworthy ADM systems, the divide between our participants notions of the socio-technical concepts of fairness, privacy, and security and their corresponding formal technical metrics suggests human oversight is insufficient on its own.
Furthermore, human conceptions of privacy, fairness, and security are contextually mutable, while ADM systems require fixed definitions to be implemented.
Thus, we, as researchers, need to develop technical definitions that align with social expectations where possible and need to stop equivocating our current technical definitions with their socio-technical counterparts. 

Altogether, current abstractions of human values for ADM systems do not sufficiently align with those impacted by them. This suggests that 
greater consideration for the effected populace is needed before any new socio-technical metrics are even proposed if we are to achieve effective alignment of expectations and accountability.
The question for policymakers, users, and implementers therefore is not just how do we design ADM systems to be human-values-centered and context-aware, but rather do the limitations of the guardrails and accountability measures required to make the system private, fair, or secure outweigh any supposed efficiencies gained by deploying the ADM system in the first place.

\section{Conclusion}
This work highlights that decision context, perceived risks and benefits, and consequences influence perceptions of decision-making approaches (HDM versus ADM) in high-stakes contexts. Participants generally preferred HDM due to concerns about accountability and the ability to handle complex situations, especially with long-term impacts on people's lives. The study found varying priorities for privacy, fairness, and robustness across different domains, indicating that acceptance of decision-making approaches is context-dependent. Moreover, participants viewed information sensitivity as dependent on usage rather than inherent data attributes. The findings suggest that fairness should be understood as a characteristic of system behavior rather than just a statistical measure, emphasizing the need for context-aware, human-centered design in AI systems for high-stakes situations. Rather than focusing only on technical metrics, AI systems should be evaluated based on their alignment with specific domain expectations and societal values.




\bibliographystyle{ACM-Reference-Format}
\bibliography{sample-base}

\appendix 

\section{Open Science} 

Our survey is provided in Appendix~\ref{appendix_questions} and Appendix~\ref{demographic_appendix}. Additional statistical analysis on human versus AI preference is provided in Appendix~\ref{quantitative_human_AI}, and participants' demographics are provided in Appendix~\ref{quantitative_human_AI}. Additionally, the detailed statistical results are in Appendix~\ref{appendix_tradeoffs_quant}. All materials provided in the appendices, as well as our scripts, are available in our Zenodo repository: \url{https://doi.org/10.5281/zenodo.19804373}

\section{Ethical Considerations}
\label{appendix_ethics}


We sought IRB approval for this study; however, we acknowledge IRB approval is insufficient and took steps beyond IRB approval to conduct our study ethically. 
When designing this study, we the authors aimed to strike a balance between asking participants about real world scenarios and causing distress to participants.
The scenarios used in this paper exist in the real world and can be the cause of real harm.
It is likely that all of our participants have interacted with ADM systems and that those systems have determined their or their loved ones' access to healthcare, housing, jobs, financial services, and/or incarceration status. 
It is also possible that our participants have been victims of unfair and unaccountable decision making at the hands of an ADM system. 

Seeking participants lived experiences about how ADM systems have affected their lives and their concerns about the risks of ADM systems was crucial to our work and we took care to present participants with fictional scenarios within real world domains to balance collecting meaningful data without causing unnecessary harm. 

Our study was guided by the principle of illuminating real, lived experiences with technology and we used an unbiased recruitment system in order to get the broadest possible variety of experiences. 
We acknowledge that by virtue of using Prolific, some experiences were excluded. 

\section{Positionality}
\label{appendix_positionality}

Our research team consists of multi-national, academic privacy and security researchers, including usability researchers, all of whom are well versed in the tradeoffs between privacy, fairness, and security. 
This undoubtedly had an effect on the survey design. 

Moreover, due to the current pervasiveness of ADM deployment, it is certain that every author has interacted with and been affected by decisions made by an ADM system. 
Our experiences with real-world ADM guided the creation of our survey, especially with regards to the different political, social, and institutional contexts of our countries of residence and origin. 
Therefore, the researchers on this study occupy a position between insiders and outsiders relative to the participants.

\section{Study Questionnaires}\label{appendix_questions}
We have eight different variations of survey questionnaires. Comprising four scenarios and two sets of tradeoffs. The survey questionnaires with fairness tradeoffs across four scenarios are presented in Appendix \ref{appendix:t1}, \ref{appendix:t2}, \ref{appendix:t3}, and \ref{appendix:t4} and the survey questionnaires with security tradeoffs across four scenarios are presented in Appendix \ref{appendix:t5}, \ref{appendix:t6}, \ref{appendix:t7}, and \ref{appendix:t8}. The demographic questions are listed in Appendix \ref{demographic_appendix}

\subsection{Fairness-Insurance}\label{appendix:t1}
InsureYou, an insurance company, sets premiums based on the risk of heart disease, charging higher premiums for high-risk patients.

\begin{enumerate}[nosep, topsep=0pt, itemsep=0pt, parsep=0pt]
    \item InsureYou relies on employees to assess an applicant’s risk of developing heart disease. 
    \begin{enumerate}[nosep, topsep=0pt, itemsep=0pt, parsep=0pt]
        \item \textbf{Imagine you are an applicant of InsureYou. How acceptable do you find it when an employee assesses the risk of developing heart disease?}
            \begin{enumerate}[nosep, topsep=0pt, itemsep=0pt, parsep=0pt]
                \item completely unacceptable 
                \item somewhat unacceptable
                \item neutral
                \item somewhat acceptable
                \item completely acceptable
            \end{enumerate}
        \item Please describe a situation where using employees would be beneficial for assessing the risk of developing heart disease.
        \item Please describe a situation where using employees would be harmful for assessing the risk of developing heart disease
    \end{enumerate}

MedForYou, a medical organization, collects data (such as age, sex, blood pressure, cholesterol, smoking habits, etc.) from patients with depression. MedForYou uses the collected data to develop an Artificial Intelligence (AI) to predict the likelihood of a patient developing a heart disease.
    \item \textbf{Instead of employees, the InsureYou now relies on MedForYou’s AI to assess an applicant’s risk of developing heart disease.}
    \begin{enumerate}[nosep, topsep=0pt, itemsep=0pt, parsep=0pt]
        \item Imagine you are an applicant of InsureYou. How acceptable do you find it when an AI assesses the risk of developing heart disease?
            \begin{enumerate}[nosep, topsep=0pt, itemsep=0pt, parsep=0pt]
                \item completely unacceptable 
                \item somewhat unacceptable
                \item neutral
                \item somewhat acceptable
                \item completely acceptable
            \end{enumerate}
        \item Please describe a situation where using AI would be beneficial for assessing the risk of developing heart disease.
        \item Please describe a situation where using AI would be harmful for assessing the risk of developing heart disease
    \end{enumerate}

    \item \textbf{Consider that InsureYou relies on MedForYou’s AI to assess an applicant’s risk of developing heart disease and calculate their insurance premiums. MedForYou identified an issue with its AI where it offers lower premiums to brunettes over non-brunettes, even if they have a high risk. To address this, MedForYou modifies the AI to mitigate the issue.}
    
    \textbullet \xspace Please describe how ‘the AI modification’ could be beneficial for everyone.
    
    \textbf{Modifying the AI has some side effects, which are described as different scenarios in parts (a), (b), (c) and (d). Please indicate your preference for the following scenarios:}
    \begin{enumerate}[nosep, topsep=0pt, itemsep=0pt, parsep=0pt]
        \item \textbf{A side effect of modifying the AI is that it incorrectly predicts higher premiums for low risk applicants. It is impossible to completely stop favoring brunettes and still provide correct insurance premiums for low risk applicants. In this scenario, if you were an applicant, which option would you suggest for MedForYou to choose?}
        \begin{enumerate}[nosep, topsep=0pt, itemsep=0pt, parsep=0pt]
            \item The AI strongly favors brunettes; otherwise, it provides the correct insurance premiums.
            \item The AI moderately favors brunettes, and it provides slightly incorrect insurance premiums.
            \item The AI slightly favors brunettes, and it provides moderately incorrect insurance premiums.
            \item The AI does not favor brunettes, and it provides highly incorrect insurance premiums
        \end{enumerate}
      \textbullet \xspace Give an example of potential issues if MedForYou chooses a different approach than your suggestion.
      \item \textbf{A side effect of modifying the AI is that InsureYou can learn whether an applicant has depression by identifying if they were part of MedForYou’s data used to create the AI. It is impossible to completely stop favoring brunettes and still prevent InsureYou from identifying an applicant has depression. In this scenario, if you were an applicant, which option would you suggest for MedForYou to choose?}
        \begin{enumerate}[nosep, topsep=0pt, itemsep=0pt, parsep=0pt]
            \item The AI strongly favors brunettes, and it has no chance to reveal whether the applicant has depression.
            \item The AI moderately favors brunettes, and it has a slight chance to reveal whether the applicant has depression.
            \item The AI slightly favors brunettes, and it has a moderate chance to reveal whether the applicant has depression.
            \item  The AI does not favor brunettes, and it has a high chance to reveal whether the applicant has depression.
        \end{enumerate}
      \textbullet \xspace Give an example of potential issues if MedForYou chooses a different approach than your suggestion.
      \item \textbf{A side effect of modifying the AI is that InsureYou can now determine the number of women in MedForYou’s data used to create the AI. It is impossible to completely stop favoring brunettes and still protect the information about the number of women. In this scenario, if you were an applicant, which option would you suggest for MedForYou to choose?}
        \begin{enumerate}[nosep, topsep=0pt, itemsep=0pt, parsep=0pt]
            \item The AI strongly favors brunettes, and it has no chance to reveal the number of women in the data.
            \item The AI moderately favors brunettes, and it has a slight chance to reveal the number of women in the data.
            \item The AI slightly favors brunettes, and it has a moderate chance to reveal the number of women in the data.
            \item The AI does not favor brunettes, and it has a high chance to reveal the number of women in the data.
        \end{enumerate}
      \textbullet \xspace Give an example of potential issues if MedForYou chooses a different approach than your suggestion.
      \item \textbf{A side effect of modifying the AI is that InsureYou’s employees could exploit the AI to disadvantage blue-eyed applicants by giving them a higher premiums. It is impossible to completely stop favoring brunettes and protect the AI from manipulation. In this scenario, if you were an applicant, which option would you suggest for MedForYou to choose?}
        \begin{enumerate}[nosep, topsep=0pt, itemsep=0pt, parsep=0pt]
            \item The AI strongly favors brunettes, and it is highly difficult to manipulate.
            \item The AI moderately favors brunettes, and it is moderately difficult to manipulate.
            \item The AI slightly favors brunettes, and it is slightly difficult to manipulate.
            \item The AI does not favor brunettes, and it is not difficult to manipulate.
        \end{enumerate}
      \textbullet \xspace Give an example of potential issues if MedForYou chooses a different approach than your suggestion.
    \end{enumerate}
        
\end{enumerate}

\subsection{Fairness-Job}\label{appendix:t2}
TechForYou, a leading technology firm, is hiring a software engineer based on candidates’ likelihood of success, favoring those with higher potential.
\begin{enumerate}[nosep, topsep=0pt, itemsep=0pt, parsep=0pt]
    \item \textbf{TechForYou relies on employees to assess an applicant’s quality.}
    \begin{enumerate}[nosep, topsep=0pt, itemsep=0pt, parsep=0pt]
        \item Imagine you are an applicant of TechForYou. How acceptable do you find it when an employee assesses the quality?
            \begin{enumerate}[nosep, topsep=0pt, itemsep=0pt, parsep=0pt]
                \item completely unacceptable 
                \item somewhat unacceptable
                \item neutral
                \item somewhat acceptable
                \item completely acceptable
            \end{enumerate}
        \item Please describe a situation where using the employee would be beneficial for assessing the quality.
        \item Please describe a situation where using the employee would be harmful for assessing the quality.
    \end{enumerate}

RecruitTalent, a recruitment firm, collects resumes (including their technical skills, academic qualifications, experience, and portfolios) from applicants with criminal backgrounds.
RecruitTalent uses the collected data to develop an Artificial Intelligence (AI) to predict whether an applicant will succeed at the job.
    \item  \textbf{Instead of employees, the TechForYou now relies on RecruitTalent’s AI to assess an applicant’s quality.}
    \begin{enumerate}[nosep, topsep=0pt, itemsep=0pt, parsep=0pt]
        \item Imagine you are an applicant of TechForYou. How acceptable do you find it when an AI assesses the quality?
            \begin{enumerate}[nosep, topsep=0pt, itemsep=0pt, parsep=0pt]
                \item completely unacceptable 
                \item somewhat unacceptable
                \item neutral
                \item somewhat acceptable
                \item completely acceptable
            \end{enumerate}
        \item Please describe a situation where using the AI would be beneficial for assessing the quality.
        \item Please describe a situation where using the AI would be harmful for assessing the quality.
    \end{enumerate}

    \item \textbf{Consider that TechForYou relies on RecruitTalent’s AI to assess an applicant’s quality and evaluate their chance of being hired. RecruitTalent identified an issue with its AI where it offers higher chances of being hired to brunettes over non-brunettes, even if they have a low potential. To address this, RecruitTalent modifies the AI to mitigate the issue.}
    
    \textbullet \xspace Please describe how ‘the AI modification’ could be beneficial for everyone.
    
   \textbf{ Modifying the AI has some side effects, which are described as different scenarios in parts (a), (b), (c) and (d). Please indicate your preference for the following scenarios:}
    \begin{enumerate}[nosep, topsep=0pt, itemsep=0pt, parsep=0pt]
        \item \textbf{A side effect of modifying the AI is that it incorrectly predicts lower chances of being hired for high potential appli- cants. It is impossible to completely stop favoring brunettes and still provide correct chance of being hired for high po- tential applicants. In this scenario, if you were an applicant, which option would you suggest for RecruitTalent to choose?}
        \begin{enumerate}[nosep, topsep=0pt, itemsep=0pt, parsep=0pt]
            \item The AI strongly favors brunettes; otherwise, it provides the correct chance of being hired.
            \item The AI moderately favors brunettes, and it provides slightly incorrect chance of being hired.
            \item The AI slightly favors brunettes, and it provides moderately incorrect chance of being hired.
            \item  The AI does not favor brunettes, and it provides highly incorrect chance of being hired
        \end{enumerate}
      \textbullet \xspace Give an example of potential issues if RecruitTalent chooses a different approach than your suggestion.
      \item \textbf{A side effect of modifying the AI is that TechForYou can learn whether an applicant has a criminal background by identifying if they were part of RecruitTalent’s data used to create the AI. It is impossible to completely stop favoring brunettes and still prevent TechForYou from identifying an applicant has a criminal background. In this scenario, if you were an applicant, which option would you suggest for RecruitTalent to choose?}
        \begin{enumerate}[nosep, topsep=0pt, itemsep=0pt, parsep=0pt]
            \item The AI strongly favors brunettes, and it has no chance to reveal whether the applicant has a criminal background.
            \item  The AI moderately favors brunettes, and it has a slight chance to reveal whether the applicant has a criminal background.
            \item The AI slightly favors brunettes, and it has a moderate chance to reveal whether the applicant has a criminal background.
            \item  The AI does not favor brunettes, and it has a high chance to reveal whether the applicant has a criminal background.
        \end{enumerate}
      \textbullet \xspace Give an example of potential issues if RecruitTalent chooses a different approach than your suggestion.
      \item \textbf{A side effect of modifying the AI is that TechForYou can now determine the number of women in RecruitTalent’s data used to create the AI. It is impossible to completely stop favoring brunettes and still protect the information about the number of women. In this scenario, if you were an applicant, which option would you suggest for RecruitTalent to choose?}
        \begin{enumerate}[nosep, topsep=0pt, itemsep=0pt, parsep=0pt]
            \item The AI strongly favors brunettes, and it has no chance to reveal the number of women in the data.
            \item The AI moderately favors brunettes, and it has a slight chance to reveal the number of women in the data.
            \item The AI slightly favors brunettes, and it has a moderate chance to reveal the number of women in the data.
            \item The AI does not favor brunettes, and it has a high chance to reveal the number of women in the data.
        \end{enumerate}
      \textbullet \xspace Give an example of potential issues if RecruitTalent chooses a different approach than your suggestion.
      \item \textbf{A side effect of modifying the AI is that TechForYou’s employees could exploit the AI to disadvantage blue-eyed applicants by giving them a lower chances of being hired. It is impossible to completely stop favoring brunettes and protect the AI from manipulation. In this scenario, if you were an applicant, which option would you suggest for Recruit- Talent to choose?}
        \begin{enumerate}[nosep, topsep=0pt, itemsep=0pt, parsep=0pt]
            \item The AI strongly favors brunettes, and it is highly difficult to manipulate.
            \item The AI moderately favors brunettes, and it is moderately difficult to manipulate.
            \item The AI slightly favors brunettes, and it is slightly difficult to manipulate.
            \item The AI does not favor brunettes, and it is not difficult to manipulate.
        \end{enumerate}
      \textbullet \xspace Give an example of potential issues if RecruitTalent chooses a different approach than your suggestion.
    \end{enumerate}
        
\end{enumerate}

\subsection{Fairness-Mortgage}\label{appendix:t3}
BankOnYourBlock, a leading financial institution, determines mortgage interest rates based on applicants’ default risk, charging higher rates for high-risk individuals.

\begin{enumerate}[nosep, topsep=0pt, itemsep=0pt, parsep=0pt]
    \item \textbf{BankOnYourBlock relies on employees to assess a client’s risk of non-payment.}
    \begin{enumerate}[nosep, topsep=0pt, itemsep=0pt, parsep=0pt]
        \item  Imagine you are a client of BankOnYourBlock. How acceptable do you find it when an employee assesses the risk of non-payment?
            \begin{enumerate}[nosep, topsep=0pt, itemsep=0pt, parsep=0pt]
                \item completely unacceptable 
                \item somewhat unacceptable
                \item neutral
                \item somewhat acceptable
                \item completely acceptable
            \end{enumerate}
        \item Please describe a situation where using employees would be beneficial for assessing the risk of non-payment.
        \item Please describe a situation where using employees would be harmful for assessing the risk of non-payment.
    \end{enumerate}

InsureYou, an insurance company, collects data on past mortgage applicants from minority demographic groups, including education, dependents, marital status, income, mortgage amount, credit history, outstanding debt, savings, investments, collateral, and employment stability. InsureYou uses the collected data to develop an Artificial Intelligence (AI) to predict whether an applicant will default.
    \item  \textbf{Instead of employees, the BankOnYourBlock now relies on InsureYou’s AI to assess a client’s risk of non-payment.}
    \begin{enumerate}[nosep, topsep=0pt, itemsep=0pt, parsep=0pt]
        \item Imagine you are a client of BankOnYourBlock. How acceptable do you find it when an AI assesses the risk of non-payment?
            \begin{enumerate}[nosep, topsep=0pt, itemsep=0pt, parsep=0pt]
                \item completely unacceptable 
                \item somewhat unacceptable
                \item neutral
                \item somewhat acceptable
                \item completely acceptable
            \end{enumerate}
        \item Please describe a situation where using AI would be beneficial for assessing the risk of non-payment.
        \item Please describe a situation where using AI would be harmful for assessing the risk of non-payment
    \end{enumerate}

    \item \textbf{Consider that BankOnYourBlock relies on InsureYou’s AI to assess a client’s risk of non-payment and calculate their interest rates. InsureYou identified an issue with its AI where it offers lower interest rates to brunettes over non-brunettes, even if they have a high risk. To address this, InsureYou modifies the AI to mitigate the issue.}
    
    \textbullet \xspace Please describe how ‘the AI modification’ could be beneficial for everyone.
    
    \textbf{Modifying the AI has some side effects, which are described as different scenarios in parts (a), (b), (c) and (d). Please indicate your preference for the following scenarios:}
    \begin{enumerate}[nosep, topsep=0pt, itemsep=0pt, parsep=0pt]
        \item \textbf{A side effect of modifying the AI is that it incorrectly predicts higher interest rates for low risk applicants. It is impossible to completely stop favoring brunettes and still pro- vide correct interest rates for low risk applicants. In this scenario, if you were a client, which option would you suggest for InsureYou to choose?}
        \begin{enumerate}[nosep, topsep=0pt, itemsep=0pt, parsep=0pt]
            \item  The AI strongly favors brunettes; otherwise, it provides the correct interest rates.
            \item The AI moderately favors brunettes, and it provides slightly incorrect interest rates.
            \item The AI slightly favors brunettes, and it provides moderately incorrect interest rates.
            \item  The AI does not favor brunettes, and it provides highly incorrect interest rates
        \end{enumerate}
      \textbullet \xspace Give an example of potential issues if InsureYou chooses a different approach than your suggestion.

      \item \textbf{A side effect of modifying the AI is that BankOnYourBlock can learn whether a client belongs to a minority demo- graphic group by identifying if they were part of InsureYou’s data used to create the AI. It is impossible to completely stop favoring brunettes and still prevent BankOnYourBlock from identifying a client belongs to a minority demographic group. In this scenario, if you were a client, which option would you suggest for InsureYou to choose?}

        \begin{enumerate}[nosep, topsep=0pt, itemsep=0pt, parsep=0pt]

            \item The AI strongly favors brunettes, and it has no chance to reveal whether the applicant belongs to a minority demographic group.

            \item   The AI moderately favors brunettes, and it has a slight chance to reveal whether the applicant belongs to a minority demographic group.

            \item The AI slightly favors brunettes, and it has a moderate chance to reveal whether the applicant belongs to a minority demographic group.

            \item  The AI does not favor brunettes, and it has a high chance to reveal whether the applicant belongs to a minority demographic group.

        \end{enumerate}
      \textbullet \xspace Give an example of potential issues if InsureYou chooses a different approach than your suggestion.

      \item \textbf{A side effect of modifying the AI is that BankOnYourBlock can now determine the number of women in InsureYou’s data used to create the AI. It is impossible to completely stop favoring brunettes and still protect the information about the number of women. In this scenario, if you were a client, which option would you suggest for InsureYou to choose?}

        \begin{enumerate}[nosep, topsep=0pt, itemsep=0pt, parsep=0pt]
            \item The AI strongly favors brunettes, and it has no chance to reveal the number of women in the data.
            \item The AI moderately favors brunettes, and it has a slight chance to reveal the number of women in the data.
            \item The AI slightly favors brunettes, and it has a moderate chance to reveal the number of women in the data.
            \item The AI does not favor brunettes, and it has a high chance to reveal the number of women in the data.
        \end{enumerate}
      \textbullet \xspace Give an example of potential issues if InsureYou chooses a different approach than your suggestion.
      \item \textbf{A side effect of modifying the AI is that BankOnYourBlock’s employees could exploit the AI to disadvantage blue-eyed applicants by giving them a higher interest rates. It is impossible to completely stop favoring brunettes and protect the AI from manipulation. In this scenario, if you were a client, which option would you suggest for InsureYou to choose?}

        \begin{enumerate}[nosep, topsep=0pt, itemsep=0pt, parsep=0pt]
            \item The AI strongly favors brunettes, and it is highly difficult to manipulate.
            \item The AI moderately favors brunettes, and it is moderately difficult to manipulate.
            \item The AI slightly favors brunettes, and it is slightly difficult to manipulate.
            \item The AI does not favor brunettes, and it is not difficult to manipulate.
        \end{enumerate}
      \textbullet \xspace Give an example of potential issues if InsureYou chooses a different approach than your suggestion.
    \end{enumerate}
        
\end{enumerate}

\subsection{Fairness-Prison}\label{appendix:t4}
A district court, CourtOnYourDistrict, determines prison sentences based on the risk of re-offending, with longer sentences for prisoners at high risk of re-offending.

\begin{enumerate}[nosep, topsep=0pt, itemsep=0pt, parsep=0pt]
    \item \textbf{CourtOnYourDistrict relies on employees to assess a prisoner’s risk of re-offending.}
    \begin{enumerate}[nosep, topsep=0pt, itemsep=0pt, parsep=0pt]
        \item   Imagine you are a prisoner of CourtOnYourDistrict. How acceptable do you find it when an employee assesses the risk of re-offending?
            \begin{enumerate}[nosep, topsep=0pt, itemsep=0pt, parsep=0pt]
                \item completely unacceptable 
                \item somewhat unacceptable
                \item neutral
                \item somewhat acceptable
                \item completely acceptable
            \end{enumerate}
        \item Please describe a situation where using employees would be beneficial for assessing the risk of re-offending.
        \item  Please describe a situation where using employees would be harmful for assessing the risk of re-offending.
    \end{enumerate}
An organization, TechForLegal, collects data on wealthy prisoners, including the nature of their offences, their criminal history, their mental health, their employability, behavioural changes, and parole officer recommendations.
TechForLegal uses the collected data to develop an Artificial Intelligence (AI) to predict whether a prisoner will re-offend.

    \item \textbf{Instead of employees, the CourtOnYourDistrict now relies on TechForLegal’s AI to assess a prisoner’s risk of re-offending.}
    \begin{enumerate}[nosep, topsep=0pt, itemsep=0pt, parsep=0pt]
        \item Imagine you are a prisoner of CourtOnYourDistrict. How acceptable do you find it when an AI assesses the risk of re-offending?
            \begin{enumerate}[nosep, topsep=0pt, itemsep=0pt, parsep=0pt]
                \item completely unacceptable 
                \item somewhat unacceptable
                \item neutral
                \item somewhat acceptable
                \item completely acceptable
            \end{enumerate}
        \item Please describe a situation where using AI would be beneficial for assessing the risk of re-offending.
        \item  Please describe a situation where using AI would be harmful for assessing the risk of re-offending.
    \end{enumerate}

    \item \textbf{Consider that CourtOnYourDistrict relies on TechForLegal’s AI to assess a prisoner’s risk of re-offending and determine their prison sentences. TechForLegal identified an issue with its AI where it offers shorter prison sentence to brunettes over non- brunettes, even if they have a high risk. To address this, TechForLegal modifies the AI to mitigate the issue.}
    
    \textbullet \xspace Please describe how ‘the AI modification’ could be beneficial for everyone.
    
    \textbf{Modifying the AI has some side effects, which are described as different scenarios in parts (a), (b), (c) and (d). Please indicate your preference for the following scenarios:}
    \begin{enumerate}[nosep, topsep=0pt, itemsep=0pt, parsep=0pt]
        \item  \textbf{A side effect of modifying the AI is that it incorrectly predicts longer prison sentence for low risk prisoners. It is impossible to completely stop favoring brunettes and still provide correct prison sentences for low risk prisoners. In this scenario, if you were a prisoner, which option would you suggest for TechForLegal to choose?}
        \begin{enumerate}[nosep, topsep=0pt, itemsep=0pt, parsep=0pt]
            \item The AI strongly favors brunettes; otherwise, it provides the correct prison sentences.
            \item The AI moderately favors brunettes, and it provides slightly incorrect prison sentences.
            \item The AI slightly favors brunettes, and it provides moderately incorrect prison sentences.
            \item  The AI does not favor brunettes, and it provides highly incorrect prison sentences
        \end{enumerate}
      \textbullet \xspace Give an example of potential issues if TechForLegal chooses a different approach than your suggestion.

      \item \textbf{A side effect of modifying the AI is that CourtOnYourDistrict can learn whether a prisoner is wealthy by identifying if they were part of TechForLegal’s data used to create the AI. It is impossible to completely stop favoring brunettes and still prevent Court-OnYourDistrict from identifying a prisoner is wealthy. In this scenario, if you were a prisoner, which option would you suggest for TechForLegal to choose?}

        \begin{enumerate}[nosep, topsep=0pt, itemsep=0pt, parsep=0pt]
            \item The AI strongly favors brunettes, and it has no chance to reveal whether the applicant is wealthy.

            \item  The AI moderately favors brunettes, and it has a slight chance to reveal whether the applicant is wealthy.

            \item  The AI slightly favors brunettes, and it has a moderate chance to reveal whether the applicant is wealthy.

            \item   The AI does not favor brunettes, and it has a high chance to reveal whether the applicant is wealthy.

        \end{enumerate}
      \textbullet \xspace Give an example of potential issues if TechForLegal chooses a different approach than your suggestion.

      \item  \textbf{A side effect of modifying the AI is that CourtOnYourDistrict can now determine the number of women in TechForLegal’s data used to create the AI. It is impossible to completely stop favoring brunettes and still protect the information about the number of women. In this scenario, if you were a prisoner, which option would you suggest for TechForLegal to choose?}

        \begin{enumerate}[nosep, topsep=0pt, itemsep=0pt, parsep=0pt]
            \item The AI strongly favors brunettes, and it has no chance to reveal the number of women in the data.
            \item The AI moderately favors brunettes, and it has a slight chance to reveal the number of women in the data.
            \item The AI slightly favors brunettes, and it has a moderate chance to reveal the number of women in the data.
            \item The AI does not favor brunettes, and it has a high chance to reveal the number of women in the data.
        \end{enumerate}
      \textbullet \xspace Give an example of potential issues if TechForLegal chooses a different approach than your suggestion.
      \item \textbf{A side effect of modifying the AI is that CourtOnYourDistrict’s employees could exploit the AI to disadvantage blue-eyed applicants by giving them a longer prison sentence. It is impossible to completely stop favoring brunettes and protect the AI from manipulation. In this scenario, if you were a prisoner, which option would you suggest for TechForLegal to choose?}

        \begin{enumerate}[nosep, topsep=0pt, itemsep=0pt, parsep=0pt]
            \item The AI strongly favors brunettes, and it is highly difficult to manipulate.
            \item The AI moderately favors brunettes, and it is moderately difficult to manipulate.
            \item The AI slightly favors brunettes, and it is slightly difficult to manipulate.
            \item The AI does not favor brunettes, and it is not difficult to manipulate.
        \end{enumerate}
      \textbullet \xspace Give an example of potential issues if TechForLegal chooses a different approach than your suggestion.
    \end{enumerate}
        
\end{enumerate}

\subsection{Security-Insurance}\label{appendix:t5}
InsureYou, an insurance company, sets premiums based on the risk of heart disease, charging higher premiums for high-risk patients.

\begin{enumerate}[nosep, topsep=0pt, itemsep=0pt, parsep=0pt]
    \item \textbf{InsureYou relies on employees to assess an applicant’s risk of developing heart disease.}
    \begin{enumerate}[nosep, topsep=0pt, itemsep=0pt, parsep=0pt]
        \item  Imagine you are an applicant of InsureYou. How acceptable do you find it when an employee assesses the risk of developing heart disease?
            \begin{enumerate}[nosep, topsep=0pt, itemsep=0pt, parsep=0pt]
                \item completely unacceptable 
                \item somewhat unacceptable
                \item neutral
                \item somewhat acceptable
                \item completely acceptable
            \end{enumerate}
        \item Please describe a situation where using employees would be beneficial for assessing the risk of developing heart disease.
        \item Please describe a situation where using employees would be harmful for assessing the risk of developing heart disease
    \end{enumerate}
MedForYou, a medical organization, collects data (such as age, sex, blood pressure, cholesterol, smoking habits, etc.) from patients with depression. MedForYou uses the collected data to develop an Artificial Intelligence (AI) to predict the likelihood of a patient developing a heart disease.

    \item \textbf{Instead of employees, the InsureYou now relies on MedForYou’s AI to assess an applicant’s risk of developing heart disease.}
    \begin{enumerate}[nosep, topsep=0pt, itemsep=0pt, parsep=0pt]
        \item  Imagine you are an applicant of InsureYou. How acceptable do you find it when an AI assesses the risk of developing heart disease?
            \begin{enumerate}[nosep, topsep=0pt, itemsep=0pt, parsep=0pt]
                \item completely unacceptable 
                \item somewhat unacceptable
                \item neutral
                \item somewhat acceptable
                \item completely acceptable
            \end{enumerate}
        \item Please describe a situation where using AI would be beneficial for assessing the risk of developing heart disease.
        \item  Please describe a situation where using AI would be harmful for assessing the risk of developing heart disease
    \end{enumerate}

    \item \textbf{Consider that InsureYou relies on MedForYou’s AI to assess an applicant’s risk of developing heart disease and calculate their insurance premiums. MedForYou identified an issue with its AI that employees could exploit the AI to disadvantage blue-eyed applicants by giving them a higher premiums, even though these applicants have a low risk. To address this, MedForYou modifies the AI to mitigate the issue.}
    
    \textbullet \xspace Please describe how ‘the AI modification’ could be beneficial for everyone.
    
    \textbf{Modifying the AI has some side effects, which are described as different scenarios in parts (a), (b), (c) and (d). Please indicate your preference for the following scenarios:}
    \begin{enumerate}[nosep, topsep=0pt, itemsep=0pt, parsep=0pt]
        \item  \textbf{A side effect of modifying the AI is that it incorrectly predicts higher premiums for low-risk applicants. It is impossible to completely protect the AI manipulation by the employee and still provide correct insurance premiums for low-risk applicants. In this scenario, if you were an applicant, which option would you suggest for MedForYou to choose?}
        \begin{enumerate}[nosep, topsep=0pt, itemsep=0pt, parsep=0pt]
            \item The AI is not highly difficult to manipulate, and it provides highly correct insurance premiums.
            \item The AI is moderately difficult to manipulate, and it provides slightly incorrect insurance premiums.
            \item The AI is slightly difficult to manipulate, and it provides moderately incorrect insurance premiums.
            \item  The AI is difficult to manipulate, and it provides highly incorrect insurance premiums.
        \end{enumerate}
      \textbullet \xspace Give an example of potential issues if MedForYou chooses a different approach than your suggestion.

      \item \textbf{A side effect of modifying the AI is that InsureYou can learn whether an applicant has depression by identifying if they were part of MedForYou’s data. It is impossible to completely protect the AI from manipulation and still prevent InsureYou from identifying applicants who have depression. In this scenario, if you were an applicant, which option would you suggest for MedForYou to choose?}

        \begin{enumerate}[nosep, topsep=0pt, itemsep=0pt, parsep=0pt]
            \item The AI is not highly difficult to manipulate, and it has no chance to reveal whether an applicant has depression.
            \item  The AI is moderately difficult to manipulate, and it has a slight chance to reveal whether an applicant has depression.
            \item  The AI is slightly difficult to manipulate, and it has a moderate chance to reveal whether an applicant has depression.
            \item    The AI is highly difficult to manipulate, and it has a high chance to reveal whether an applicant has depression.
        \end{enumerate}
        
      \textbullet \xspace Give an example of potential issues if MedForYou chooses a different approach than your suggestion.

      \item  \textbf{The above modification of the AI has made it possible for InsureYou to determine the number of women in MedForYou’s collected data. It is impossible to completely protect the AI from manipulation and protect information about the number of women. In this scenario, if you were an applicant, which option would you suggest for MedForYou to choose?}

        \begin{enumerate}[nosep, topsep=0pt, itemsep=0pt, parsep=0pt]
            \item The AI is not highly difficult to manipulate, and it has no chance to reveal the number of women in the data.
            \item The AI is moderately difficult to manipulate, and it has a slight chance to reveal the number of women in the data.
            \item The AI is slightly difficult to manipulate, and it has a moderate chance to reveal the number of women in the data.
            \item  The AI is highly difficult to manipulate, and it has high chance to reveal the number of women in the data.
        \end{enumerate}
        
      \textbullet \xspace Give an example of potential issues if MedForYou chooses a different approach than your suggestion.
      
      \item \textbf{A side effect of modifying the AI is that InsureYou’s employees can easily learn the information of the participants that MedForYou used to build the AI. It is impossible to completely protect the AI from manipulation while preventing information learning. In this scenario, if you were an applicant, which option would you suggest for MedForYou to choose?}

        \begin{enumerate}[nosep, topsep=0pt, itemsep=0pt, parsep=0pt]
            \item The AI is not highly difficult to manipulate, and there is no chance of learning the information used to build the AI.
            \item The AI is moderately difficult to manipulate, and there is a slight chance to learn the information used to build the AI.
            \item The AI is slightly difficult to manipulate, and there is a moderate chance to learn the information used to build the AI.
            \item The AI is difficult to manipulate, and there is a high chance to learn the information used to build the AI.
        \end{enumerate}
        
      \textbullet \xspace Give an example of potential issues if MedForYou chooses a different approach than your suggestion.
    \end{enumerate}
        
\end{enumerate}

\subsection{Security-Job}\label{appendix:t6}
TechForYou, a leading technology firm, is hiring a software engineer based on candidates’ likelihood of success, favoring those with higher potential.

\begin{enumerate}[nosep, topsep=0pt, itemsep=0pt, parsep=0pt]
    \item \textbf{TechForYou relies on employees to assess an applicant’s quality.}
    \begin{enumerate}[nosep, topsep=0pt, itemsep=0pt, parsep=0pt]
        \item  Imagine you are an applicant of TechForYou. How acceptable do you find it when an employee assesses the quality?
            \begin{enumerate}[nosep, topsep=0pt, itemsep=0pt, parsep=0pt]
                \item completely unacceptable 
                \item somewhat unacceptable
                \item neutral
                \item somewhat acceptable
                \item completely acceptable
            \end{enumerate}
        \item Please describe a situation where using employees would be beneficial for assessing the quality.
        \item Please describe a situation where using employees would be harmful for assessing the quality.
    \end{enumerate}
    
RecruitTalent, a recruitment firm, collects resumes (including their technical skills, academic qualifications, experience, and portfolios) from applicants with criminal backgrounds.
RecruitTalent uses the collected data to develop an Artificial Intelligence (AI) to predict whether an applicant will succeed at the job.

    \item \textbf{ Instead of employees, the TechForYou now relies on RecruitTalent’s AI to assess an applicant’s quality.}
    \begin{enumerate}[nosep, topsep=0pt, itemsep=0pt, parsep=0pt]
        \item  Imagine you are an applicant of TechForYou. How acceptable do you find it when an AI assesses the quality?
            \begin{enumerate}[nosep, topsep=0pt, itemsep=0pt, parsep=0pt]
                \item completely unacceptable 
                \item somewhat unacceptable
                \item neutral
                \item somewhat acceptable
                \item completely acceptable
            \end{enumerate}
        \item Please describe a situation where using AI would be beneficial for assessing the quality.
        \item Please describe a situation where using AI would be harmful for assessing the quality.
    \end{enumerate}

    \item \textbf{Consider that TechForYou relies on RecruitTalent’s AI to assess an applicant’s quality and evaluate their chance of being hired. RecruitTalent identified an issue with its AI that employees could exploit the AI to disadvantage blue-eyed applicants by giving them a lower chance of being hired, even though these applicants have a high potential. To address this, RecruitTalent modifies the AI to mitigate the issue.}
    
    \textbullet \xspace Please describe how ‘the AI modification’ could be beneficial for everyone.
    
    \textbf{Modifying the AI has some side effects, which are described as different scenarios in parts (a), (b), (c) and (d). Please indicate your preference for the following scenarios:}
    \begin{enumerate}[nosep, topsep=0pt, itemsep=0pt, parsep=0pt]
        \item  \textbf{A side effect of modifying the AI is that it incorrectly predicts lower chances of being hired for high-potential applicants. It is impossible to completely protect the AI manipulation by the employee and still provide a correct chance of being hired for high-potential applicants. In this scenario, if you were an applicant, which option would you suggest for RecruitTalent to choose?}
        \begin{enumerate}[nosep, topsep=0pt, itemsep=0pt, parsep=0pt]
            \item The AI is not highly difficult to manipulate, and it provides a correct chance of being hired.
            \item The AI is moderately difficult to manipulate, and it provides a slightly incorrect chance of being hired.
            \item The AI is slightly difficult to manipulate, and it provides a moderately incorrect chance of being hired.
            \item The AI is highly difficult to manipulate, and it provides a highly incorrect chance of being hired.
        \end{enumerate}
      \textbullet \xspace Give an example of potential issues if RecruitTalent chooses a different approach than your suggestion.

      \item \textbf{A side effect of modifying the AI is that TechForYou can learn whether an applicant has a criminal background by identifying if they were part of RecruitTalent’s data. It is impossible to completely protect the AI from manipulation and still prevent TechForYou from identifying applicants has a criminal background. In this scenario, if you were an applicant, which option would you suggest for RecruitTalent to choose?}

        \begin{enumerate}[nosep, topsep=0pt, itemsep=0pt, parsep=0pt]
           \item The AI is not difficult to manipulate, and it has no chance to reveal whether an applicant has a criminal background.
            \item  The AI is moderately difficult to manipulate, and it has a slight chance to reveal whether an applicant has a criminal background.
            \item  The AI is slightly difficult to manipulate, and it has a moderate chance to reveal whether an applicant has a criminal background.
            \item  The AI is highly difficult to manipulate, and it has high chance to reveal whether an applicant has a criminal background.
        \end{enumerate}
        
      \textbullet \xspace Give an example of potential issues if RecruitTalent chooses a different approach than your suggestion.

      \item  \textbf{The above modification of the AI has made it possible for TechForYou to determine the number of women in RecruitTalent’s collected data. It is impossible to completely protect the AI from manipulation and protect information about the number of women. In this scenario, if you were an applicant, which option would you suggest for RecruitTalent to choose?}

        \begin{enumerate}[nosep, topsep=0pt, itemsep=0pt, parsep=0pt]
            \item The AI is not highly difficult to manipulate, and it has no chance to reveal the number of women in the data.
            \item The AI is moderately difficult to manipulate, and it has a  slight chance to reveal the number of women in the data.
            \item The AI is slightly difficult to manipulate, and it has a moderate chance to reveal the number of women in the data.
            \item  The AI is highly difficult to manipulate, and it has high chance to reveal the number of women in the data.
        \end{enumerate}
        
      \textbullet \xspace Give an example of potential issues if RecruitTalent chooses a different approach than your suggestion.
      
      \item \textbf{A side effect of modifying the AI is that TechForYou’s employees can easily learn the information of the participants that RecruitTalent used to build the AI. It is impossible to completely protect the AI from manipulation while preventing information learning. In this scenario, if you were an applicant, which option would you suggest for RecruitTalent to choose?}

        \begin{enumerate}[nosep, topsep=0pt, itemsep=0pt, parsep=0pt]
            \item The AI is not difficult to manipulate, and there is no chance to learn the information used to build the AI.
            \item The AI is moderately difficult to manipulate, and there is a  slight chance to learn the information used to build the AI.
            \item The AI is slightly difficult to manipulate, and there is a moderate chance to learn the information used to build the AI.
            \item The AI is highly difficult to manipulate, and there is high chance to learn the information used to build the AI.
        \end{enumerate}
        
      \textbullet \xspace Give an example of potential issues if RecruitTalent chooses a different approach than your suggestion.
    \end{enumerate}
        
\end{enumerate}

\subsection{Security-Mortgage}\label{appendix:t7}
BankOnYourBlock, a leading financial institution, determines mortgage interest rates based on applicants’ default risk, charging higher rates for high-risk individuals.

\begin{enumerate}[nosep, topsep=0pt, itemsep=0pt, parsep=0pt]
    \item \textbf{BankOnYourBlock relies on employees to assess a client’s risk of non-payment.}
    \begin{enumerate}[nosep, topsep=0pt, itemsep=0pt, parsep=0pt]
        \item  Imagine you are a client of BankOnYourBlock. How acceptable do you find it when an employee assesses the risk of non-payment?
            \begin{enumerate}[nosep, topsep=0pt, itemsep=0pt, parsep=0pt]
                \item completely unacceptable 
                \item somewhat unacceptable
                \item neutral
                \item somewhat acceptable
                \item completely acceptable
            \end{enumerate}
        \item  Please describe a situation where using employees would be beneficial for assessing the risk of non-payment.
        \item Please describe a situation where using employees would be harmful for assessing the risk of non-payment

    \end{enumerate}
    
InsureYou, an insurance company, collects data on past mortgage applicants from minority demographic groups, including education, dependents, marital status, income, mortgage amount, credit history, outstanding debt, savings, investments, collateral, and employment stability. InsureYou uses the collected data to develop an Artificial Intelligence (AI) to predict whether an applicant will default.

    \item \textbf{Instead of employees, the BankOnYourBlock now relies on InsureYou’s AI to assess a client’s risk of non-payment.}
    \begin{enumerate}[nosep, topsep=0pt, itemsep=0pt, parsep=0pt]
        \item Imagine you are a client of BankOnYourBlock. How acceptable do you find it when an AI assesses the risk of non-payment?
            \begin{enumerate}[nosep, topsep=0pt, itemsep=0pt, parsep=0pt]
                \item completely unacceptable 
                \item somewhat unacceptable
                \item neutral
                \item somewhat acceptable
                \item completely acceptable
            \end{enumerate}
        \item  Please describe a situation where using AI would be beneficial for assessing the risk of non-payment.
        \item Please describe a situation where using AI would be harmful for assessing the risk of non-payment

    \end{enumerate}

    \item \textbf{Consider that BankOnYourBlock relies on InsureYou’s AI to assess an applicant’s risk of non-payment and calculate their interest rates. InsureYou identified an issue with its AI that employees could exploit the AI to disadvantage blue-eyed applicants by giving them a higher interest rate, even though these applicants have a low risk. To address this, InsureYou modifies the AI to mitigate the issue.}
    
    \textbullet \xspace Please describe how ‘the AI modification’ could be beneficial for everyone.
    
    \textbf{Modifying the AI has some side effects, which are described as different scenarios in parts (a), (b), (c) and (d). Please indicate your preference for the following scenarios:}
    \begin{enumerate}[nosep, topsep=0pt, itemsep=0pt, parsep=0pt]
        \item  \textbf{A side effect of modifying the AI is that it incorrectly predicts higher interest rates for low-risk applicants. It is impossible to completely protect the AI manipulation by the employee and still provide correct interest rates for low-risk applicants. In this scenario, if you were a client, which option would you suggest for InsureYou to choose?}
        \begin{enumerate}[nosep, topsep=0pt, itemsep=0pt, parsep=0pt]
            \item The AI is not difficult to manipulate, and it provides correct interest rates.
            \item The AI is moderately difficult to manipulate, and it provides slightly incorrect interest rates.
            \item The AI is slightly difficult to manipulate, and it provides moderately incorrect interest rates.
            \item  The AI is highly difficult to manipulate, and it provides highly incorrect interest rates.
        \end{enumerate}
      \textbullet \xspace Give an example of potential issues if InsureYou chooses a different approach than your suggestion.

      \item \textbf{A side effect of modifying the AI is that BankOnYourBlock can learn whether a client belongs to a minority demographic group by identifying if they were part of InsureYou’s data. It is impossible to completely protect the AI from manipulation and still prevent BankOnYourBlock from identifying applicants belongs to a minority demographic group. In this scenario, if you were a client, which option would you suggest for InsureYou to choose?}

        \begin{enumerate}[nosep, topsep=0pt, itemsep=0pt, parsep=0pt]
            \item The AI is not difficult to manipulate, and it has no chance to reveal whether a client belongs to a minority demographic group.
            \item The AI is moderately difficult to manipulate, and it has a slight chance to reveal whether a client belongs to a minority demographic group.
            \item The AI is slightly difficult to manipulate, and it has a moderate chance to reveal whether a client belongs to a minority demographic group.
            \item  The AI is highly difficult to manipulate, and it has a high chance to reveal whether a client belongs to a minority demographic group.
        \end{enumerate}
        
      \textbullet \xspace Give an example of potential issues if InsureYou chooses a different approach than your suggestion.

      \item  \textbf{The above modification of the AI has made it possible for BankOnYourBlock to determine the number of women in InsureYou’s collected data. It is impossible to completely protect the AI from manipulation and protect information about the number of women. In this scenario, if you were a client, which option would you suggest for InsureYou to choose?}

        \begin{enumerate}[nosep, topsep=0pt, itemsep=0pt, parsep=0pt]
            \item The AI is not difficult to manipulate, and it has no chance to reveal the number of women in the data.
            \item The AI is moderately difficult to manipulate, and it has a slight chance to reveal the number of women in the data.
            \item The AI is slightly difficult to manipulate, and it has a moderate chance to reveal the number of women in the data.
            \item  The AI is highly difficult to manipulate, and it has high chance to reveal the number of women in the data.
        \end{enumerate}
        
      \textbullet \xspace Give an example of potential issues if InsureYou chooses a different approach than your suggestion.
      
      \item \textbf{A side effect of modifying the AI is that BankOnYourBlock’s employees can easily learn the information of the participants that InsureYou used to build the AI. It is impossible to completely protect the AI from manipulation while preventing information learning. In this scenario, if you were a client, which option would you suggest for InsureYou to choose?}

        \begin{enumerate}[nosep, topsep=0pt, itemsep=0pt, parsep=0pt]
            \item The AI is not difficult to manipulate, and there is no chance to learn the information used to build the AI.
            \item The AI is moderately difficult to manipulate, and there is a slight chance to learn the information used to build the AI.
            \item The AI is slightly difficult to manipulate, and there is a moderate chance to learn the information used to build the AI.
            \item The AI is highly difficult to manipulate, and there is no chance to learn the information used to build the AI.
        \end{enumerate}
        
      \textbullet \xspace Give an example of potential issues if InsureYou chooses a different approach than your suggestion.
    \end{enumerate}
        
\end{enumerate}

\subsection{Security-Prison}\label{appendix:t8}
A district court, CourtOnYourDistrict, determines prison sentences based on the risk of re-offending, with longer sentences for prisoners at high risk of re-offending.

\begin{enumerate}[nosep, topsep=0pt, itemsep=0pt, parsep=0pt]
    \item \textbf{CourtOnYourDistrict relies on employees to assess a prisoner’s risk of re-offending.}
    \begin{enumerate}[nosep, topsep=0pt, itemsep=0pt, parsep=0pt]
        \item  Imagine you are a prisoner of CourtOnYourDistrict. How acceptable do you find it when an employee assesses the risk of re-offending?
            \begin{enumerate}[nosep, topsep=0pt, itemsep=0pt, parsep=0pt]
                \item completely unacceptable 
                \item somewhat unacceptable
                \item neutral
                \item somewhat acceptable
                \item completely acceptable
            \end{enumerate}
        \item  Please describe a situation where using employees would be beneficial for assessing the risk of re-offending.
        \item  Please describe a situation where using employees would be harmful for assessing the risk of re-offending.

    \end{enumerate}
    
An organization, TechForLegal, collects data on wealthy prisoners, including the nature of their offences, their criminal history, their mental health, their employability, behavioural changes, and parole officer recommendations.
TechForLegal uses the collected data to develop an Artificial Intelligence (AI) to predict whether a prisoner will re-offend.

    \item \textbf{ Instead of employees, the CourtOnYourDistrict now relies on TechForLegal’s AI to assess a prisoner’s risk of re-offending.}
    \begin{enumerate}[nosep, topsep=0pt, itemsep=0pt, parsep=0pt]
        \item Imagine you are a prisoner of CourtOnYourDistrict. How acceptable do you find it when an AI assesses the risk of re-offending?

            \begin{enumerate}[nosep, topsep=0pt, itemsep=0pt, parsep=0pt]
                \item completely unacceptable 
                \item somewhat unacceptable
                \item neutral
                \item somewhat acceptable
                \item completely acceptable
            \end{enumerate}
        \item  Please describe a situation where using AI would be beneficial for assessing the risk of re-offending.
        \item  Please describe a situation where using AI would be harmful for assessing the risk of re-offending.

    \end{enumerate}

    \item \textbf{Consider that CourtOnYourDistrict relies on TechForLegal’s AI to assess an applicant’s risk of re-offending and determine their prison sentences. TechForLegal identified an issue with its AI that employees could exploit the AI to disadvantage blue- eyed prisoners by giving them a longer prison sentence, even though these prisoners have a low risk. To address this, TechForLegal modifies the AI to mitigate the issue.
}
    
    \textbullet \xspace Please describe how ‘the AI modification’ could be beneficial for everyone.
    
    \textbf{Modifying the AI has some side effects, which are described as different scenarios in parts (a), (b), (c) and (d). Please indicate your preference for the following scenarios:}
    \begin{enumerate}[nosep, topsep=0pt, itemsep=0pt, parsep=0pt]
        \item  \textbf{A side effect of modifying the AI is that it incorrectly predicts longer prison sentence for low risk prisoners. It is impossible to completely protect the AI manipulation by the employee and still provide correct prison sentences for low risk prisoners. In this scenario, if you were a prisoner, which option would you suggest for TechForLegal to choose?}
        \begin{enumerate}[nosep, topsep=0pt, itemsep=0pt, parsep=0pt]
            \item The AI is not difficult to manipulate, and it provides correct prison sentences.
            \item The AI is moderately difficult to manipulate, and it provides slightly incorrect prison sentences.
            \item The AI is slightly difficult to manipulate, and it provides moderately incorrect prison sentences.
            \item  The AI is highly difficult to manipulate, and it provides highly incorrect prison sentences.

        \end{enumerate}
      \textbullet \xspace Give an example of potential issues if TechForLegal chooses a different approach than your suggestion.

      \item \textbf{A side effect of modifying the AI is that CourtOnYourDistrict can learn whether a prisoner is wealthy by identifying if they were part of TechForLegal’s data. It is impossible to completely protect the AI from manipulation and still prevent CourtOnYourDistrict from identifying prisoners is wealthy. In this scenario, if you were a prisoner, which option would you suggest for TechForLegal to choose?}

        \begin{enumerate}[nosep, topsep=0pt, itemsep=0pt, parsep=0pt]
            \item The AI is not difficult to manipulate, and it has a high chance to reveal whether the prisoner is wealthy.
            \item The AI is moderately difficult to manipulate, and it has a slight chance to reveal whether the prisoner is wealthy.
            \item The AI is slightly difficult to manipulate, and it has a moderate chance to reveal whether the prisoner is wealthy.
            \item  The AI is highly difficult to manipulate, and it has a high chance to reveal whether the prisoner is wealthy.
        \end{enumerate}
        
      \textbullet \xspace Give an example of potential issues if TechForLegal chooses a different approach than your suggestion.

      \item  \textbf{The above modification of the AI has made it possible for CourtOnYourDistrict to determine the number of women in TechForLegal’s collected data. It is impossible to completely protect the AI from manipulation and protect information about the number of women. In this scenario, if you were a prisoner, which option would you suggest for TechForLegal to choose?}

        \begin{enumerate}[nosep, topsep=0pt, itemsep=0pt, parsep=0pt]
            \item The AI is not difficult to manipulate, and it has no chance to reveal the number of women in the data.
            \item The AI is moderately difficult to manipulate, and it has a slight chance to reveal the number of women in the data.
            \item The AI is slightly difficult to manipulate, and it has a moderate chance to reveal the number of women in the data.
            \item  The AI is highly difficult to manipulate, and it has a high chance to reveal the number of women in the data.
        \end{enumerate}
        
      \textbullet \xspace Give an example of potential issues if TechForLegal chooses a different approach than your suggestion.
      
      \item \textbf{A side effect of modifying the AI is that CourtOnYourDistrict’s employees can easily learn the information of the patticipants that TechForLegal used to build the AI. It is impossible to completely protect the AI from manipulation while preventing information learning. In this scenario, if you were a prisoner, which option would you suggest for TechForLegal to choose?}

        \begin{enumerate}[nosep, topsep=0pt, itemsep=0pt, parsep=0pt]
            \item The AI is not highly difficult to manipulate, and there is no chance to learn the information used to build the AI.
            \item The AI is moderately difficult to manipulate, and there is a slight chance to learn the information used to build the AI.
            \item The AI is slightly difficult to manipulate, and there is a moderate chance to learn the information used to build the AI.
            \item The AI is highly difficult to manipulate, and there is a high chance to learn the information used to build the AI.
        \end{enumerate}
        
      \textbullet \xspace Give an example of potential issues if TechForLegal chooses a different approach than your suggestion.
    \end{enumerate}
        
\end{enumerate}

\section{Demographic}
\label{demographic_appendix}

    \begin{enumerate}[nosep, topsep=0pt, itemsep=0pt, parsep=0pt]
        \item \textbf{What is your gender?}\\
        $\bullet$ Man \quad $\bullet$ Woman \quad $\bullet$ Non-Binary \quad $\bullet$ Prefer not to answer \quad $\bullet$ Self-describe

        \item \textbf{Do you identify as being part of any minority groups?} \\
        $\bullet$ Yes \quad $\bullet$ No \quad $\bullet$ Prefer not to answer

        \item \textbf{What age group do you belong to?} \\
        $\bullet$ 18--25 \quad $\bullet$ 26--35 \quad $\bullet$ 36--45 \quad $\bullet$ 46--55 \quad $\bullet$ 56--65 \quad $\bullet$ Prefer not to answer

        \item \textbf{What is your highest level of education?} \\
        $\bullet$ Less than a high school degree \\
        $\bullet$ High school degree or equivalent \\
        $\bullet$ Non-university education beyond high school (technical college, community college, vocational education) \\
        $\bullet$ Some college, no degree \\
        $\bullet$ Bachelor’s or Associate’s degree \\
        $\bullet$ Master’s degree \\
        $\bullet$ Doctoral degree \\
        $\bullet$ Prefer not to answer

        \item \textbf{Are you a practitioner (researcher, working professional) in Machine Learning?} \\
        $\bullet$ Yes \quad $\bullet$ No \quad $\bullet$ Prefer not to answer
    \end{enumerate}

\subsection{Demographic Distribution}
Tables \ref{tab:gender_distribution_human_ai}, \ref{tab:minority_distribution_human_ai}, \ref{tab:practitioner_distribution_human_ai} show the gender distribution, minority group distribution, and ML practitioner distribution, respectively. 
\begin{table}[!htbp]
\centering
\caption{Participant gender distribution in human and AI decision-making preference across scenarios. \\} 
\label{tab:gender_distribution_human_ai}
\begin{tabular}{@{}lccc@{}}
\toprule
\textbf{Scenario} & \textbf{N} & \textbf{Men} & \textbf{Women} \\ \midrule
Insurance   & 191 & 89 & 100 \\
Job         & 192 & 95 & 95  \\
Mortgage    & 197 & 91 & 102 \\
Prison     & 193 & 94 & 93 \\
 \bottomrule
\end{tabular}

\end{table}

\begin{table}[!htbp]
\centering
\caption{Minority group distribution in human and AI decision-making preferences across scenarios, excluding participants who preferred not to answer.}
\vspace{.5em}
\label{tab:minority_distribution_human_ai}
\begin{tabular}{@{}lcccc@{}}
\toprule
\textbf{Scenario} &\textbf{N}  & \textbf{Minority} & \textbf{Non-Minority} \\
\midrule
Insurance   & 185 & 85 & 100 \\
Job         & 186 & 70 & 116 \\
Mortgage    & 191 & 96 & 95  \\
Prison      & 188 & 78 & 110 \\
\bottomrule
\end{tabular}

\end{table}

\begin{table}[!htbp]
\centering
\caption{Distribution of ML practitioners’ preferences for human versus AI decision-making across scenarios, excluding participants who preferred not to answer.}
\vspace{.5em}
\label{tab:practitioner_distribution_human_ai}
\begin{tabular}{@{}lcccc@{}}

\toprule
\textbf{Scenario} & \textbf{N} & \textbf{Practitioner} & \textbf{Non-Practitioner}    \\ \midrule
Insurance   & 192 & 32 & 158 \\
Job         & 192 & 25 & 165\\
Mortgage    & 197 & 23 & 168 \\
Prison      & 198 & 13 & 176 \\
 \bottomrule
\end{tabular}

\end{table}

\section{Human versus ADM: Quantitative Analysis}\label{quantitative_human_AI}
We provide additional details as tables for our statistical results presented in the main body of the work.

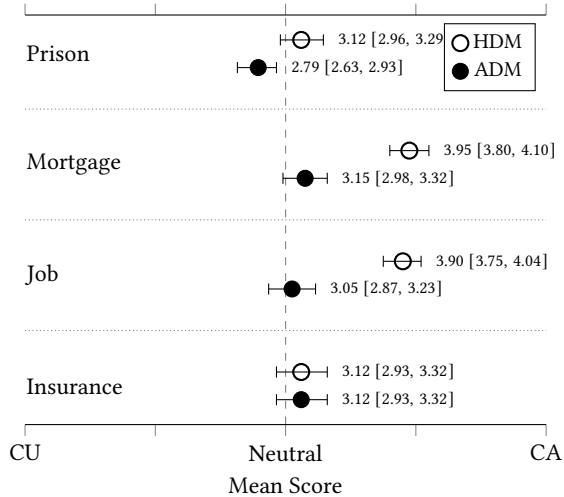
\begin{figure}[!htbp]
\centering
\caption{Mean scores and 95\% confidence intervals (CI) for the acceptability of HDM and ADM across scenarios. Acceptability is measured on a five-point scale, from completely unacceptable (CU) to completely acceptable (CA).}
\vspace{.8em}
\label{fig:mean_ci}
\begin{tikzpicture}
\begin{axis}[
    width=\columnwidth,
    height=7.0cm,
    xmin=1,
    xmax=5,
    ymin=0.8,
    ymax=8.2,
    xtick={1,2,3,4,5},
    axis line style={draw=none}, 
    xticklabels={CU,
    , 
    Neutral, 
    , 
    CA},
    xlabel={Mean Score},
    ytick={7.5,5.5,3.5,1.5},
    yticklabels={
        Prison,
        Mortgage,
        Job,
        Insurance
    },
     every y tick label/.style={
        anchor=near yticklabel opposite,
        xshift=0.2em,
    },   
    ytick style={draw=none}, 
    tick align=outside, 
    clip=false,
    legend style={
        at={(0.98,0.98)},
        anchor=north east,
        draw=black,
        fill=white,
        font=\small
    }
]
\addplot[
    gray,
    dashed,
    forget plot
]
coordinates {(3,0.8) (3,8.2)};
\addplot[
    gray,
    densely dotted,
    forget plot
]
coordinates {(1,6.5) (5,6.5)};

\addplot[
    gray,
    densely dotted,
    forget plot
]
coordinates {(1,4.5) (5,4.5)};
\addplot[
    gray,
    densely dotted,
    forget plot
]
coordinates {(1,2.5) (5,2.5)};
\draw
    (axis cs:2.96,7.75) -- (axis cs:3.29,7.75);
\draw
    (axis cs:2.96,7.65) -- (axis cs:2.96,7.85);
\draw
    (axis cs:3.29,7.65) -- (axis cs:3.29,7.85);
\draw
    (axis cs:2.63,7.25) -- (axis cs:2.93,7.25);
\draw
    (axis cs:2.63,7.15) -- (axis cs:2.63,7.35);
\draw
    (axis cs:2.93,7.15) -- (axis cs:2.93,7.35);
\draw
    (axis cs:3.80,5.75) -- (axis cs:4.10,5.75);
\draw
    (axis cs:3.80,5.65) -- (axis cs:3.80,5.85);
\draw
    (axis cs:4.10,5.65) -- (axis cs:4.10,5.85);
\draw
    (axis cs:2.98,5.25) -- (axis cs:3.32,5.25);
\draw
    (axis cs:2.98,5.15) -- (axis cs:2.98,5.35);
\draw
    (axis cs:3.32,5.15) -- (axis cs:3.32,5.35);
\draw
    (axis cs:3.75,3.75) -- (axis cs:4.04,3.75);
\draw
    (axis cs:3.75,3.65) -- (axis cs:3.75,3.85);
\draw
    (axis cs:4.04,3.65) -- (axis cs:4.04,3.85);
\draw
    (axis cs:2.87,3.25) -- (axis cs:3.23,3.25);
\draw
    (axis cs:2.87,3.15) -- (axis cs:2.87,3.35);
\draw
    (axis cs:3.23,3.15) -- (axis cs:3.23,3.35);
\draw
    (axis cs:2.93,1.75) -- (axis cs:3.32,1.75);
\draw
    (axis cs:2.93,1.65) -- (axis cs:2.93,1.85);
\draw
    (axis cs:3.32,1.65) -- (axis cs:3.32,1.85);
\draw
    (axis cs:2.93,1.25) -- (axis cs:3.32,1.25);
\draw
    (axis cs:2.93,1.15) -- (axis cs:2.93,1.35);
\draw
    (axis cs:3.32,1.15) -- (axis cs:3.32,1.35);


\addplot[
    only marks,
    mark=o,
    mark size=3pt,
    black,
    thick
]
coordinates {
    (3.12,7.75)
    (3.95,5.75)
    (3.90,3.75)
    (3.12,1.75)
};
\addlegendentry{HDM}

\addplot[
    only marks,
    mark=*,
    mark size=3pt,
    black
]
coordinates {
    (2.79,7.25)
    (3.15,5.25)
    (3.05,3.25)
    (3.12,1.25)
};
\addlegendentry{ADM}


\node[
    anchor=west,
    font=\scriptsize
] at (axis cs:3.34,7.75)
{$3.12\ [2.96,\,3.29]$};

\node[
    anchor=west,
    font=\scriptsize
] at (axis cs:2.98,7.25)
{$2.79\ [2.63,\,2.93]$};

\node[
    anchor=west,
    font=\scriptsize
] at (axis cs:4.15,5.75)
{$3.95\ [3.80,\,4.10]$};
\node[
    anchor=west,
    font=\scriptsize
] at (axis cs:3.37,5.25)
{$3.15\ [2.98,\,3.32]$};
\node[
    anchor=west,
    font=\scriptsize
] at (axis cs:4.09,3.75)
{$3.90\ [3.75,\,4.04]$};

\node[
    anchor=west,
    font=\scriptsize
] at (axis cs:3.28,3.25)
{$3.05\ [2.87,\,3.23]$};

\node[
    anchor=west,
    font=\scriptsize
] at (axis cs:3.37,1.75)
{$3.12\ [2.93,\,3.32]$};

\node[
    anchor=west,
    font=\scriptsize
] at (axis cs:3.37,1.25)
{$3.12\ [2.93,\,3.32]$};

\draw (rel axis cs:0,0) -- (rel axis cs:1,0);
\draw (rel axis cs:0,1) -- (rel axis cs:1,1);

\end{axis}
\end{tikzpicture}

\end{figure}

\addd{Figure \ref{fig:mean_ci} shows the visual representation of the mean and confidence intervals of the HDM and ADM acceptability.}

Table \ref{wilcoxon:human:ai} includes the summary of the Wilcoxon signed-rank test, showing that except for the insurance scenario, all the other scenarios showed a significant difference in acceptability of HDM versus ADM and achieved $p < 0.05$. 
Table \ref{tab:accept_human} and \ref{tab:accept_ai} show the six pairwise comparisons across the scenarios for HDM and ADM, respectively. 


\begin{table*}[!htbp]
\centering
\caption{Pairwise comparisons of scenarios for HDM acceptability across scenarios. Each row tests the null hypothesis that the scenario one and scenario two distributions are the same. Asymptotic significances (2-sided tests) are displayed. The significance level is $.050$.}
\begin{tabularx}{\textwidth}{l *{4}{>{\centering\arraybackslash}X}}
\toprule

\multirow{2}{*}{\textbf{Scenario Pair}} 
    & \textbf{Test} 
    & \textbf{Standard} 
    & \textbf{Significance} 
    & \textbf{Adjusted} \\

    & \textbf{Statistic} 
    & \textbf{Test Statistic} 
    & \textbf{(p-value)} 
    & \textbf{p-value} \\

\hline
 Prison - Insurance      & 13.590	&.620	&.535	&1.000\\
Prison - Job            &137.587	&6.273	& \textbf{$<.001$}	&\textbf{.000}\\
Prison - Mortgage       &150.404		&6.902	& \textbf{$<.001$}	&\textbf{.000}\\
Insurance - Job          &-123.997		&-5.632	& \textbf{$<.001$}	&\textbf{.000} \\
Insurance - Mortgage      & -136.815		 &-6.254	 & \textbf{$<.001$}	 &\textbf{.000}\\
 Job - Mortgage         &-12.817	&-.586	&.558	&1.000\\
\hline
\end{tabularx}
\label{tab:accept_human}
\end{table*}

\begin{table}[H]
\centering
\caption{Summary of the Wilcoxon signed-rank tests comparing human and AI acceptability across four decision-making contexts, evaluated at a significance level of $0.005$.}
\label{wilcoxon:human:ai}
\vspace{.5em}
\renewcommand{\arraystretch}{1.3}
\begin{tabularx}{\linewidth}{lcccc}
\toprule
\textbf{Scenario} & \textbf{N} & \textbf{Z} & \textbf{p-value} & \textbf{Interpretation}\\
\midrule
Prison & 195  & $-2.873$ & \boldmath$.004$ & Significant\\

Mortgage & 197  & $-6.904$ & \boldmath$<.001$ & Significant\\

Job & 192  & $-6.794$ & \boldmath$<.001$ & Significant\\

Insurance & 192  & $-0.253$ & 0.800 & Not significant\\

\bottomrule

\end{tabularx}
\end{table}

\begin{table*}[!htbp]
\centering
\caption{Pairwise comparisons of scenarios for ADM acceptability across scenarios. Each row tests the null hypothesis that the scenario one and scenario two distributions are the same. Asymptotic significances (2-sided tests) are displayed. The significance level is $.050$.}
\begin{tabularx}{\textwidth}{l *{4}{>{\centering\arraybackslash}X}}
\toprule

\multirow{2}{*}{\textbf{Scenario Pair}} 
    & \textbf{Test} 
    & \textbf{Standard} 
    & \textbf{Significance} 
    & \textbf{Adjusted} \\

    & \textbf{Statistic} 
    & \textbf{Test Statistic} 
    & \textbf{(p-value)} 
    & \textbf{p-value} \\
\hline
Prison - Insurance      &59.030		&2.672	&.008	&.045\\
Prison - Job            &46.883		&2.125	&.034	&\textbf{.202}\\
Prison - Mortgage       &63.605		&2.901	&.004	&\textbf{.022}\\
 Insurance - Job          &12.147	&.548	&.584	&1.000\\
 Insurance - Mortgage      &-4.575   &-.208	&.836	&1.000\\
 Job - Mortgage         &-16.722	 &-.760	 &.447	 &1.000\\
\hline
\end{tabularx}
\label{tab:accept_ai}
\end{table*}





\section{Tradeoff Preferences: Across Scenarios}\label{appendix_tradeoffs_quant}


\begin{table}[!htbp]
\centering
\caption{Chi-Square test result for participants' preference distribution for the eight tradeoffs across the scenarios.  The significance
level is .05.}
\begin{tabular}{lccc}
\hline
\textbf{Tradeoff} & \textbf{$p-value$} & \textbf{$\chi^2$}\\
\hline
Fairness vs. Accuracy & $.700$ & $6.391$ \\
Fairness vs. Membership Inference     & \boldmath{$ <.001$} & $.377$  \\
Fairness vs. Distribution Inference  & \boldmath{$<.001$} & $16.969$\\
Fairness vs. Security    & $.076$ & $15.568$. \\
Security vs. Accuracy    & $.119$ & $14.095$ \\
Security vs. Membership Inference     & $034$  & $18.121$\\
Security vs. Distribution Inference     & $632$  & $7.045$\\
Security vs. Data Reconstruction     & $.195$  & $12.344$ \\
\hline
\end{tabular}
\label{tab:crosstab_all_tradeoffs}
\end{table}

\addd{Table \ref{tab:crosstab_all_tradeoffs} shows that participants did not show any significant difference in preference across the scenarios except the tradeoffs, fairness versus membership inference, and fairness versus distribution inference. Further, we present the chi-square test results for scenario pairs where we found significant differences in preference within a tradeoff. In the tradeoff fairness versus membership inference and fairness versus distribution inference, we have found significant differences in preference.   Tables \ref{tab:crosstab_tradeoff2} and \ref{tab:crosstab_tradeoff3} show the scenario pairs that cause the significant differences. }
\begin{table}[H]
\centering
\caption{Chi-Square test result for the pairwise comparisons of scenarios for the tradeoff between fairness and membership inference. The significance
level is .0083.}
\begin{tabular}{lccc}
\hline
\textbf{Scenarios} & \textbf{$p-value$} & \textbf{$\chi^2$}\\
\hline
Insurance - Job & \boldmath{$<.001$} & 37.944 \\
Insurance - Mortgage     & $0.23$ & $9.529$  \\
Insurance - Prison  & \boldmath{$<.001$} & $16.969$\\
Job - Mortgage    & \boldmath{$<.001$} & $29.103$  \\
Job - Prison    & $.155$ & $5.240$  \\
Mortgage - Prison    & \boldmath$.008$  & $11.881$\\
\hline
\end{tabular}
\label{tab:crosstab_tradeoff2}
\end{table}

\begin{table}[H]
\centering
\caption{Chi-Square test result for the pairwise comparisons of scenarios for the tradeoff between fairness and distribution inference. The significance
level is .0083.}
\begin{tabular}{lccc}
\hline
\textbf{Scenarios} & \textbf{$p-value$} & \textbf{$\chi^2$}\\
\hline
Insurance - Job & \boldmath{$.002$} & $14.817$ \\
Insurance - Mortgage     & $.549$ & $2.117$  \\
Insurance - Prison  & \boldmath{$<.001$} & $27.800$\\
Job - Mortgage    & \boldmath{$.005$} & $12.862$  \\
Job - Prison    & $.390$ & $3.010$  \\
Mortgage -. Prison    & \boldmath{$<.001$}  & $22.110$\\
\hline
\end{tabular}
\label{tab:crosstab_tradeoff3}
\end{table}

\section{Tradeoff Preferences: Within Scenarios}\label{appendix_tradeoffs_within_scenarios}
In this section, we present the chi-square test results for the eight tradeoffs. When we find a significant difference in a tradeoff, we further perform chi-square residual analysis to determine which option was chosen significantly more or less than the others across the four scenarios.  
\subsection{Tradeoff: Fairness versus accuracy}\label{appendix_tradeoff1}

Table \ref{tab:tradeoff1_chi_suare} shows that, across all scenarios, participants showed a significant preference in choosing among the four response categories. Table 
\ref{tab:tradeoff1_chi_square_prison},
\ref{tab:tradeoff1_chi_square_mortgage}, \ref{tab:tradeoff1_chi_square_insurance} show that, in scenarios prison, mortgage, and insurance, participants significantly preferred ``High bias; No Inaccuracy". In scenario job, participants did not significantly prefer one option but we found that they significantly less selected the option ``Slight bias; Moderate Inaccuracy", shown in Table \ref{tab:tradeoff1_chi_square_job}.

\begin{table}[H]
\centering
\caption{Tradeoff: Fairness versus accuracy Chi-square test summary. The significance level is .05.}
\vspace{.5em}
\label{tab:tradeoff1_chi_suare}
\begin{tabular}{@{}lcccc@{}}
\toprule
\textbf{Scenario} & \textbf{Chi-Square} & \textbf{df} & \textbf{p-value} \\
\midrule

Prison      & 39.626 & 3 & \boldmath$<.001$  \\
Mortgage    & 28.567 & 3 & \boldmath$<.001$ \\
Job         & 15.061 & 3 & \boldmath$.002$ \\
Insurance   & 34.175 & 3 & \boldmath$<.001$ \\

 \bottomrule
\end{tabular}
\end{table}

\begin{table*}[!htbp]
\centering
\caption{Chi-Square residual analysis for prison response categories for tradeoff fairness versus accuracy.   
Standardized residuals were compared against a Bonferroni-adjusted critical value of $z=\pm2.50$, corresponding to $\alpha=.0125$ for four category-level comparisons.}
\vspace{.5em}
\label{tab:tradeoff1_chi_square_prison}
\begin{tabular}{@{}lcccc@{}}
\toprule
\textbf{Category} & \textbf{Observed $N$} & \textbf{Expected $N$} & \textbf{Residual} & \textbf{Standardized Residual $z$} \\
\midrule

High bias; No Inaccuracy & 45 & 24.8 & 20.3 & \textbf{4.08} \\
Moderate bias; Slight Inaccuracy & 35 & 24.8 & 10.3 & 2.07 \\
Slight bias; Moderate Inaccuracy & 9 & 24.8 & -15.8 & \textbf{-3.17} \\
No bias; High Inaccuracy & 10 & 24.8 & -14.8 & \textbf{-2.97} \\
\bottomrule
\end{tabular}
\end{table*}

\begin{table*}[!htbp]
\centering
\caption{Chi-Square residual analysis for mortgage response categories for tradeoff fairness versus accuracy.   
Standardized residuals were compared against a Bonferroni-adjusted critical value of $z=\pm2.50$, corresponding to $\alpha=.0125$ for four category-level comparisons.}
\vspace{.5em}
\label{tab:tradeoff1_chi_square_mortgage}
\begin{tabular}{@{}lcccc@{}}
\toprule
\textbf{Category} & \textbf{Observed $N$} & \textbf{Expected $N$} & \textbf{Residual} & \textbf{Standardized Residual $z$} \\
\midrule

High bias; No Inaccuracy & 42 & 24.3 & 17.8 & \textbf{3.61} \\
Moderate bias; Slight Inaccuracy & 31 & 24.3 & 6.8 & 1.38 \\
Slight bias; Moderate Inaccuracy & 8 & 24.3 & -16.3 & \textbf{-3.31} \\
No bias; High Inaccuracy & 16 & 24.3 & -8.3 & -1.68 \\
\bottomrule
\end{tabular}
\end{table*}

\begin{table*}[!htbp]
\centering
\caption{Chi-Square residual analysis for job response categories for tradeoff fairness versus accuracy.   
Standardized residuals were compared against a Bonferroni-adjusted critical value of $z=\pm2.50$, corresponding to $\alpha=.0125$ for four category-level comparisons.}
\vspace{.5em}
\label{tab:tradeoff1_chi_square_job}
\begin{tabular}{@{}lcccc@{}}
\toprule
\textbf{Category} & \textbf{Observed $N$} & \textbf{Expected $N$} & \textbf{Residual} & \textbf{Standardized Residual $z$} \\
\midrule
High bias; No Inaccuracy & 37 & 24.8 & 12.3 & 2.47 \\
Moderate bias; Slight Inaccuracy & 31 & 24.8 & 6.3 & 1.27 \\
Slight bias; Moderate Inaccuracy & 13 & 24.8 & -11.8 & \textbf{-2.37} \\
No bias; High Inaccuracy & 18 & 24.8 & -6.8 & -1.37 \\
\bottomrule
\end{tabular}

\end{table*}

\begin{table*}[!htbp]
\centering
\caption{Chi-Square residual analysis for insurance response categories for tradeoff fairness versus accuracy.   
Standardized residuals were compared against a Bonferroni-adjusted critical value of $z= \pm2.50$, corresponding to $α=.0125$ for four category-level comparisons.}
\vspace{.5em}
\label{tab:tradeoff1_chi_square_insurance}
\begin{tabular}{@{}lcccccc@{}}
\toprule
\textbf{Category} & \textbf{Observed $N$} & \textbf{Expected $N$} & \textbf{Residual}  & \textbf{Standardized Residual $z$} \\
\midrule

High bias; No Inaccuracy & 44 & 24.3 & 19.8 & \textbf{4.02} \\
Moderate bias; Slight Inaccuracy   & 32 & 24.3 & 7.8       & 1.58 \\
Slight bias; Moderate Inaccuracy  & 11 & 24.3 & -13.3    & \textbf{-2.70}  \\
No bias; High Inaccuracy & 10 & 24.3 & -14.3      & \textbf{-2.90}  \\
 \bottomrule
\end{tabular}

\end{table*}

\begin{table}[H]
\centering
\caption{Tradeoff: Fairness versus membership inference Chi-square test summary.  The significance level is .05.}
\vspace{.5em}
\label{tab:tradeoff2_chi_suare}
\begin{tabular}{@{}lcccc@{}}
\toprule
\textbf{Scenario} & \textbf{Chi-Square} & \textbf{df} & \textbf{p-value} \\
\midrule
Prison      & 30.091 & 3 & \boldmath$<.001$  \\
Mortgage    & 13.309 & 3 & \boldmath$.004$ \\
Job         & 78.899 & 3 & \boldmath$<.001$ \\
Insurance   & 4.649 & 3 & $.199$ \\

 \bottomrule
\end{tabular}

\end{table}

\subsection{Tradeoff: Fairness versus  membership inference}\label{appendix_tradeoff2}

Table \ref{tab:tradeoff2_chi_suare} shows that, across all scenarios except insurance, participants showed a significant preference in choosing among the four response categories. Table \ref{tab:tradeoff2_chi_square_prison} and  \ref{tab:tradeoff2_chi_square_job} show that in the prison and job scenarios, participants significantly preferred the ``No bias; High chance of MI" category. However, in the mortgage scenario, participants significantly less preferred ``Slight bias; moderate chance of MI", shown in Table  \ref{tab:tradeoff2_chi_square_mortgage}.

\begin{table*}[!htbp]
\centering
\caption{Chi-Square residual analysis for prison response categories for tradeoff fairness versus membership inference. Standardized residuals were compared against a Bonferroni-adjusted critical value of $z=\pm2.50$, corresponding to $\alpha=.0125$ for four category-level comparisons.}
\vspace{.5em}
\label{tab:tradeoff2_chi_square_prison}
\begin{tabular}{@{}lcccc@{}}
\toprule
\textbf{Category} & \textbf{Observed $N$} & \textbf{Expected $N$} & \textbf{Residual} & \textbf{Standardized Residual $z$} \\
\midrule

High bias; No chance of MI & 21 & 24.8 & -3.8 & -0.76 \\
Moderate bias; Slight chance of MI & 15 & 24.8 & -9.8 & -1.97 \\
Slight bias; Moderate chance of MI & 15 & 24.8 & -9.8 & -1.97  \\
No bias; High chance of MI & 48 & 24.8 & 23.3 & \textbf{4.68} \\
\bottomrule
\end{tabular}
\end{table*}

\begin{table*}[!htbp]
\centering
\caption{Chi-Square residual analysis for mortgage response categories for tradeoff fairness versus membership inference.   
Standardized residuals were compared against a Bonferroni-adjusted critical value of $z=\pm2.50$, corresponding to $\alpha=.0125$ for four category-level comparisons.}
\vspace{.5em}
\label{tab:tradeoff2_chi_square_mortgage}
\begin{tabular}{@{}lcccc@{}}
\toprule
\textbf{Category} & \textbf{Observed $N$} & \textbf{Expected $N$} & \textbf{Residual} & \textbf{Standardized Residual $z$} \\
\midrule

High bias; No chance of MI & 32 & 24.3 & 7.8   & 1.58  \\
Moderate bias; Slight chance of MI & 27 & 24.3 & 2.8   & 0.57 \\
Slight bias; Moderate chance of MI & 9  & 24.3 & -15.3 & \textbf{-3.10} \\
No bias; High chance of MI  & 29 & 24.3 & 4.8   & .97  \\
\bottomrule
\end{tabular}
\end{table*}

\begin{table*}[!htbp]
\centering
\caption{Chi-Square residual analysis for job response categories for tradeoff fairness versus membership inference (MI).   
Standardized residuals were compared against a Bonferroni-adjusted critical value of $z=\pm2.50$, corresponding to $\alpha=.0125$ for four category-level comparisons.}
\vspace{.5em}
\label{tab:tradeoff2_chi_square_job}
\begin{tabular}{@{}lcccc@{}}
\toprule
\textbf{Category} & \textbf{Observed $N$} & \textbf{Expected $N$} & \textbf{Residual} & \textbf{Standardized Residual $z$} \\
\midrule

High bias; No chance of MI & 12 & 24.8 & -12.8 & \textbf{-2.57}  \\
Moderate bias; Slight chance of MI & 11 & 24.8 & -13.8 & \textbf{-2.77} \\
Slight bias; Moderate chance of MI & 13 & 24.8 & -11.8 & \textbf{-2.37} \\
No bias; High chance of MI & 63 & 24.8 & 38.3  & \textbf{7.69} \\
\bottomrule
\end{tabular}

\end{table*}

\subsection{Tradeoff: Fairness versus  distribution inference}\label{appendix_tradeoff3}

Table \ref{tab:tradeoff3_chi_suare} shows that, in the scenario, job and prison participants showed a significant preference in choosing among the four response categories. Table \ref{tab:tradeoff3_chi_square_prison} and  \ref{tab:tradeoff3_chi_square_job} show that participants significantly preferred the highest level of fairness, which is ``No bias; High chance of DI".

\begin{table}[H]
\centering
\caption{Tradeoff: Fairness versus distribution inference (DI) Chi-square test summary.  The significance level is .05.}
\vspace{.5em}
\label{tab:tradeoff3_chi_suare}
\begin{tabular}{@{}lcccc@{}}
\toprule
\textbf{Scenario} & \textbf{Chi-Square} & \textbf{df} & \textbf{p-value} \\
\midrule

Insurance   & 3.990 & 3 & $.263$ \\
Job         & 40.030 & 3 & \boldmath$<.001$ \\
Mortgage    & 10.583 & 3 & $.014$ \\
Prison      & 79.384 & 3 & \boldmath$<.001$  \\
 \bottomrule
\end{tabular}

\end{table}

\begin{table*}[!htbp]
\centering
\caption{Chi-Square residual analysis for prison response categories for tradeoff fairness versus distribution inference.   
Standardized residuals were compared against a Bonferroni-adjusted critical value of $z=\pm2.50$, corresponding to $\alpha=.0125$ for four category-level comparisons.}
\vspace{.5em}
\label{tab:tradeoff3_chi_square_prison}
\begin{tabular}{@{}lcccc@{}}
\toprule
\textbf{Category} & \textbf{Observed $N$} & \textbf{Expected $N$} & \textbf{Residual} & \textbf{Standardized Residual $z$} \\
\midrule

High bias; No chance of DI  & 16 & 24.8 & -8.8 & -1.97 \\
Moderate bias; Slight chance of DI& 16 & 24.8 & -8.8 & \textbf{-2.77} \\
Slight bias; Moderate chance of DI  & 15 & 24.8 & -9.8 & \textbf{-2.97} \\
No bias; High chance of DI & 52 & 24.8 & 27.3 & \textbf{7.69}  \\
\bottomrule
\end{tabular}
\end{table*}

\begin{table*}[!htbp]
\centering
\caption{Chi-Square residual analysis for job response categories for tradeoff fairness versus distribution inference.   
Standardized residuals were compared against a Bonferroni-adjusted critical value of $z=\pm2.50$, corresponding to $\alpha=.0125$ for four category-level comparisons.}
\vspace{.5em}
\label{tab:tradeoff3_chi_square_job}
\begin{tabular}{@{}lcccc@{}}
\toprule
\textbf{Category} & \textbf{Observed $N$} & \textbf{Expected $N$} & \textbf{Residual} & \textbf{Standardized Residual $z$} \\
\midrule

High bias; No chance of DI  & 16 & 24.8 & -8.8 & -1.77 \\
Moderate bias; Slight chance of DI  & 16 & 24.8 & -8.8 & -1.77  \\
Slight bias; Moderate chance of DI   & 15 & 24.8 & -9.8 & -1.97  \\
No bias; High chance of DI  & 52 & 24.8 & 27.3 & \textbf{5.48} \\
\bottomrule
\end{tabular}

\end{table*}

\subsection{Tradeoff: Fairness versus  security}\label{appendix_tradeoff4}
Table~\ref{tab:tradeoff4_chi_suare}, which shows that, in mortgage and prison scenarios, 
participants showed a significant preference in choosing among the four response categories. Table \ref{tab:tradeoff4_chi_square_prison} and  \ref{tab:tradeoff4_chi_square_mortgage} show that, across the four categories, participants significantly preferred the highest level of protection against adversarial tampering, which is ``High bias; No Manipulation Risk".

\begin{table}[H]
\centering
\caption{Tradeoff: Fairness versus security  Chi-square test summary.  The significance level is .05.}
\vspace{.5em}
\label{tab:tradeoff4_chi_suare}
\begin{tabular}{@{}lcccc@{}}
\toprule
\textbf{Scenario} & \textbf{Chi-Square} & \textbf{df} & \textbf{p-value} \\
\midrule

Insurance   & 6.511 & 3 & $.089$ \\
Job         & 5.525 & 3 & $.137$ \\
Mortgage    & 18.083 & 3 & \boldmath$<.001$ \\
Prison      & 24.800 & 3 & \boldmath$<.001$  \\
 \bottomrule
\end{tabular}

\end{table}

\begin{table*}[!htbp]
\centering
\caption{Chi-Square residual analysis for prison response categories for tradeoff fairness versus security.   
Standardized residuals were compared against a Bonferroni-adjusted critical value of $z=\pm2.50$, corresponding to $\alpha=.0125$ for four category-level comparisons.}
\vspace{.5em}
\label{tab:tradeoff4_chi_square_prison}
\begin{tabular}{@{}lcccc@{}}
\toprule
\textbf{Category} & \textbf{Observed $N$} & \textbf{Expected $N$} & \textbf{Residual} & \textbf{Standardized Residual $z$} \\
\midrule

High bias; No Manipulation Risk  & 46 & 25.0 & 21.0  & \textbf{4.20} \\
Slight bias; Moderate Manipulation Risk  & 14 & 25.0 & -11.0 & -2.20 \\
Moderate bias; Slight Manipulation Risk & 22 & 25.0 & -3.0  & -0.60 \\
No Bias; High Manipulation Risk  & 18 & 25.0 & -7.0  & -1.40\\
\bottomrule
\end{tabular}
\end{table*}

\begin{table*}[!htbp]
\centering
\caption{Chi-Square residual analysis for mortgage response categories for tradeoff fairness versus security.   
Standardized residuals were compared against a Bonferroni-adjusted critical value of $z=\pm2.50$, corresponding to $\alpha=.0125$ for four category-level comparisons.}
\vspace{.5em}
\label{tab:tradeoff4_chi_square_mortgage}
\begin{tabular}{@{}lcccc@{}}
\toprule
\textbf{Category} & \textbf{Observed $N$} & \textbf{Expected $N$} & \textbf{Residual} & \textbf{Standardized Residual $z$} \\
\midrule

High bias; No Manipulation Risk  & 41 & 24.0 & 17.0 & \textbf{3.47} \\
Slight bias; Moderate Manipulation Risk  & 24 & 24.0 & 0.0  & 0.00 \\
Moderate bias; Slight Manipulation Risk & 16 & 24.0 & -8.0 & -1.63 \\
No Bias; High Manipulation Risk & 15 & 24.0 & -9.0 & -1.84 \\
\bottomrule
\end{tabular}
\end{table*}

\subsection{Tradeoff: Security versus accuracy}\label{appendix_tradeoff5}

Table \ref{tab:tradeoff5_chi_suare} shows that, across all scenarios except prison, participants showed a significant preference in choosing among the four response categories. In the mortgage and job scenarios, participants significantly chose ``Moderate manipulation risk; Slight Inaccuracy" over the other categories, as shown in Tables \ref{tab:tradeoff5_chi_square_job} and \ref{tab:tradeoff5_chi_square_mortgage}. In the insurance scenario, participants chose the accuracy-maximizing option, as shown in Figure  \ref{tab:tradeoff5_chi_square_insurance}.

\begin{table}[H]
\centering
\caption{Tradeoff: security versus accuracy Chi-square test summary.  The significance level is .05.}
\vspace{.5em}
\label{tab:tradeoff5_chi_suare}
\begin{tabular}{@{}lcccc@{}}

\toprule
\textbf{Scenario} & \textbf{Chi-Square} & \textbf{df} & \textbf{p-value} \\
\midrule

Insurance   & 15.872 & 3 & \boldmath$.001$ \\
Job         & 23.688 & 3 & \boldmath$<.001$ \\
Mortgage    & 23.760 & 3 & \boldmath$<.001$ \\
Prison      & 5.842 & 3 & $.120$  \\
 \bottomrule
\end{tabular}
\end{table}

\begin{table*}[!htbp]
\centering
\caption{Chi-Square residual analysis for mortgage response categories for tradeoff security versus accuracy.   
Standardized residuals were compared against a Bonferroni-adjusted critical value of $z=\pm2.50$, corresponding to $\alpha=.0125$ for four category-level comparisons.}
\vspace{.5em}
\label{tab:tradeoff5_chi_square_mortgage}
\begin{tabular}{@{}lcccc@{}}
\toprule
\textbf{Category} & \textbf{Observed $N$} & \textbf{Expected $N$} & \textbf{Residual} & \textbf{Standardized Residual $z$} \\
\midrule

High manipulation risk;
No Inaccuracy & 34 & 25.0 & 9.0   & 1.80\\
Moderate manipulation risk;
Slight Inaccuracy & 40 & 25.0 & 15.0  & \textbf{3.00} \\
Slight manipulation risk;
Moderate Inaccuracy & 13 & 25.0 & -12.0 & -2.40 \\
No manipulation risk;
High Inaccuracy & 13 & 25.0 & -12.0 & -2.40 \\
\bottomrule
\end{tabular}
\end{table*}

\begin{table*}[!htbp]
\centering
\caption{Chi-Square residual analysis for job response categories for tradeoff security versus accuracy.   
Standardized residuals were compared against a Bonferroni-adjusted critical value of $z=\pm2.50$, corresponding to $\alpha=.0125$ for four category-level comparisons.}
\vspace{.5em}
\label{tab:tradeoff5_chi_square_job}
\begin{tabular}{@{}lcccc@{}}
\toprule
\textbf{Category} & \textbf{Observed $N$} & \textbf{Expected $N$} & \textbf{Residual} & \textbf{Standardized Residual $z$} \\
\midrule

High manipulation risk;
No Inaccuracy & 26 & 23.3 & 2.8   & 0.58 \\
Moderate manipulation risk;
Slight Inaccuracy & 41 & 23.3 & 17.8  & \textbf{3.69} \\
Slight manipulation risk;
Moderate Inaccuracy & 16 & 23.3 & -7.3  & -1.51 \\
No manipulation risk;
High Inaccuracy & 10 & 23.3 & -13.3 & \textbf{-2.76} \\

\bottomrule
\end{tabular}

\end{table*}

\begin{table*}[!htbp]
\centering
\caption{Chi-Square residual analysis for insurance response categories for tradeoff security versus accuracy.   
Standardized residuals were compared against a Bonferroni-adjusted critical value of $z= \pm2.50$, corresponding to $α=.0125$ for four category-level comparisons.}
\vspace{.5em}
\label{tab:tradeoff5_chi_square_insurance}
\begin{tabular}{@{}lcccccc@{}}

\toprule
\textbf{Category} & \textbf{Observed $N$} & \textbf{Expected $N$} & \textbf{Residual}  & \textbf{Standardized Residual $z$} \\
\midrule

High manipulation risk; No Inaccuracy & 36 & 23.5 & 12.5  & \textbf{2.58} \\
Moderate manipulation risk;
Slight Inaccuracy & 29 & 23.5 & 5.5   & 1.13 \\
Slight manipulation risk;
Moderate Inaccuracy & 18 & 23.5 & -5.5  & -1.31 \\
No manipulation risk;
High Inaccuracy & 11 & 23.5 & -12.5 & \textbf{-2.58} \\
 \bottomrule
\end{tabular}

\end{table*}

\subsection{Tradeoff: Security versus membership inference}\label{appendix_tradeoff6}

Table~\ref{tab:tradeoff6_chi_suare} shows that, across all scenarios except mortgage, participants did not show a significant difference in preference in choosing among the four response categories.
In mortgage, participants chose ``Slight manipulation risk;
Moderate chance of MI" significantly less than the other three options, shown in Table~\ref{tab:tradeoff6_chi_square_mortgage}.

\begin{table}[H]
\centering
\caption{Tradeoff: security versus membership inference (MI) Chi-square test summary.  The significance level is .05.}
\vspace{.5em}
\label{tab:tradeoff6_chi_suare}
\begin{tabular}{@{}lcccc@{}}

\toprule
\textbf{Scenario} & \textbf{Chi-Square} & \textbf{df} & \textbf{p-value} \\
\midrule

Insurance   & 2.426 & 3 & $.489$ \\
Job         & 8.806 & 3 & $.032$ \\
Mortgage    & 14.160 & 3 & \boldmath$.003$ \\
Prison      & 1.379 & 3 & $.710$  \\
 \bottomrule
\end{tabular}
\end{table}

\begin{table*}[!htbp]
\centering
\caption{Chi-Square residual analysis for mortgage response categories for tradeoff security versus membership inference.  Standardized residuals were compared against a Bonferroni-adjusted critical value of $z=\pm2.50$, corresponding to $\alpha=.0125$ for four category-level comparisons.}
\vspace{.5em}
\label{tab:tradeoff6_chi_square_mortgage}
\begin{tabular}{@{}lcccc@{}}
\toprule
\textbf{Category} & \textbf{Observed $N$} & \textbf{Expected $N$} & \textbf{Residual} & \textbf{Standardized Residual $z$} \\
\midrule

High manipulation risk;
No chance of MI & 30 & 25.0 & 5.0   & 1.00 \\
Moderate manipulation risk;
Slight chance of MI & 37 & 25.0 & 12.0  & 2.40 \\
Slight manipulation risk;
Moderate chance of MI & 12 & 25.0 & -13.0 & \textbf{-2.60} \\
No manipulation risk;
High chance of MI & 21 & 25.0 & -4.0  & -0.80 \\
\bottomrule
\end{tabular}
\end{table*}

\begin{table}[H]
\centering
\caption{Tradeoff: security versus distribution inference (DI) Chi-square test summary.  The significance level is .05.}
\vspace{.5em}
\label{tab:tradeoff7_chi_square}
\begin{tabular}{@{}lcccc@{}}

\toprule
\textbf{Scenario} & \textbf{Chi-Square} & \textbf{df} & \textbf{p-value} \\
\midrule

Insurance   & 8.200 & 3 & $.042$ \\
Job         & 1.151 & 3 & $.765$ \\
Mortgage    & 1.360 & 3 & $.715$ \\
Prison      & 5.574 & 3 & $.134$  \\
 \bottomrule
\end{tabular}
\end{table}

\subsection{Tradeoff: Security versus distribution inference}\label{appendix_tradeoff7}
Table \ref{tab:tradeoff7_chi_square} shows that, across all scenarios, participants did not choose any option significantly more or less than the others.

\begin{table*}[htbp]
\centering
\caption{Chi-Square residual analysis for mortgage response categories for tradeoff security versus data reconstruction.   
Standardized residuals were compared against a Bonferroni-adjusted critical value of $z=\pm2.50$, corresponding to $\alpha=.0125$ for four category-level comparisons.}
\vspace{.5em}
\label{tab:tradeoff8_chi_square_mortgage}
\begin{tabular}{@{}lcccc@{}}
\toprule
\textbf{Category} & \textbf{Observed $N$} & \textbf{Expected $N$} & \textbf{Residual} & \textbf{Standardized Residual $z$} \\
\midrule

High manipulation risk;
No chance of DR  & 28 & 25.0 & 3.0   & 0.69 \\
Moderate manipulation risk;
Slight chance of DR & 34 & 25.0 & 9.0   & 2.08 \\
Slight manipulation risk;
Moderate chance of DR & 9  & 25.0 & -16.0 & \textbf{-3.20} \\
No manipulation risk;
High chance of DR & 29 & 25.0 & 4.0   & 0.92 \\
\bottomrule
\end{tabular}
\end{table*}

\subsection{Tradeoff: Security versus data reconstruction}\label{appendix_tradeoff8}

Table \ref{tab:tradeoff8_chi_suare} shows that, except for mortgage, across all scenarios, participants did not choose any option significantly more or less than the others. In mortgage, participants chose ``Slight manipulation risk; Moderate chance of DR" significantly less than the others, shown in Table \ref{tab:tradeoff8_chi_square_mortgage}.

\begin{table}[H]
\centering
\caption{Tradeoff: security versus data reconstruction (DR) Chi-square test summary.  The significance level is .05.}
\vspace{.5em}
\label{tab:tradeoff8_chi_suare}
\begin{tabular}{@{}lcccc@{}}

\toprule
\textbf{Scenario} & \textbf{Chi-Square} & \textbf{df} & \textbf{p-value} \\
\midrule

Insurance   & 5.319 & 3 & $.150$ \\
Job         & 4.957 & 3 & $.175$ \\
Mortgage    & 14.480 & 3 & \boldmath$.002$ \\
Prison      & 7.000 & 3 & $.072$  \\
 \bottomrule
\end{tabular}
\end{table}


\end{document}